\def\sqr#1#2{{\vcenter{\vbox{\hrule height.#2pt\hbox{\vrule width.#2pt 
height#1pt \kern#1pt \vrule width.#2pt}\hrule height.#2pt}}}}
\def\d{\partial}

\def\w{\mathchoice\sqr45\sqr45\sqr{2.1}3\sqr{1.5}3\,} 
\def\=d{\,{\buildrel\rm def\over =}\,}

\documentclass{article}
\usepackage{amssymb}\usepackage{amsmath}
\newcommand{\CC}{\mathbb C}
\newcommand{\RR}{\mathbb R}

\newcommand{\NN}{\mathbb N}
\newcommand{\supp}{{\rm supp\>}}
\begin{document}
\title{Algebraic Quantum Field Theory, Perturbation Theory,
and the Loop Expansion}
\author{M. D\"utsch\thanks{Work supported by the Deutsche 
Forschungsgemeinschaft}  and K. Fredenhagen\\[1mm]
Institut f\"ur Theoretische Physik\\
Universit\"at Hamburg\\
Luruper Chaussee 149\\
D-22761 Hamburg, Germany\\
{\tt duetsch@mail.desy.de, fredenha@x4u2.desy.de}}

\date{}

\maketitle

{\bf Dedicated to the memory of Harry Lehmann}\\
\\
\begin{abstract}
The perturbative treatment of quantum field theory is formulated 
within the framework of algebraic quantum field theory. 
We show that the algebra of interacting fields is additive, i.e. fully
determined by its subalgebras associated to arbitrary small subregions
of Minkowski space.   
We also give an algebraic formulation of the loop
expansion by introducing a projective system ${\cal A}^{(n)}$ of
observables ``up to $n$ loops'' where ${\cal A}^{(0)}$ is the Poisson algebra
of the classical field theory. Finally we give a local 
algebraic formulation for two cases of the quantum action principle and
compare it with the usual formulation in terms of Green's functions.


\end{abstract}

\section{Introduction}
Quantum field theory is a very successful frame for our present understanding
of elementary particle physics. In the case of QED it led to fantastically
precise predictions of experimentally measurable quantities; moreover 
the present standard model of elementary particle physics is of a similar
structure and is also in good agreement with experiments. Unfortunately, 
it is not so clear what an interacting quantum field theory really is, 
expressed in meaningful mathematical terms. 
In particular, it is by no means evident how the local algebras of 
observables can be defined. A direct approach by methods of constructive 
field theory led to the paradoxical conjecture that QED does not exist; 
the situation seems to be better for Yang-Mills theories because of 
asymptotic freedom, but there the problem of big fields which can appear
at large volumes poses at present unsurmountable problems \cite{Bal,MRS}.

In this paper we will take a pragmatic point of view: 
interacting quantum field theory certainly exists on
the level of perturbation theory, and our confidence on quantum field theory
relies mainly on the agreement of experimental data with results from low
orders of perturbation theory. On the other hand, the general structure of 
algebraic quantum field theory (or 'local quantum physics') coincides
nicely with the qualitative features of elementary particle physics, 
therefore it seems to be worthwhile to revisit perturbation theory from 
the point of view of algebraic quantum field theory. This will, on the 
one hand side, provide physically relevant examples for algebraic 
quantum field theory, and on the other hand, give new
insight into the structure of perturbation theory. In particular, we will see,
that we can reach a complete separation of the infrared problem from the
ultraviolet problem. This might be of relevance for Yang-Mills theory, and
it is important for the construction of the theory on curved 
spacetimes \cite{BF}.

The plan of the paper is as follows. We will start by describing the
St\"uckelberg-Bogoliubov-Shirkov-Epstein-Glaser-version of perturbation
theory \cite{BS,EG,St2,S,BF}. This construction yields the local
$S$-matrices $S(g)\>(g\in {\cal D}(\RR^4))$ as formal power series 
in $g$ (Sect. 2). The most important requirement  
which is used in this construction is 
the condition of causality (\ref{2.12}) which is a functional equation
for $g\rightarrow S(g)$. The results of Sects. 3 and 4 
are to a large extent valid beyond 
perturbation theory. We only assume that
we are given a family of unitary solutions of the condition of 
causality. In terms of these local $S$-matrices
we will construct nets of local observable
algebras for the interacting  theory (sect. 3).  
We will see that, as a consequence of causality, the interacting 
theory is completely determined if it is known for arbitrary small 
spacetime volumes (Sect. 4).

In Sect. 5 we algebraically quantize a free field by deforming 
the (classical) Poisson algebra. In a second step we generalize
this quantization procedure to the perturbative interacting
field. We end up with an algebraic formulation of the expansion in 
$\hbar$ of the interacting observables ('loop expansion').

In the last section we investigate two examples for the quantum action 
principle: the field equation and the variation of a parameter in the 
interaction. Usually this principle is formulated in terms of Green's
functions \cite{L,Lam,PS}, i.e. the vacuum expectation values 
of timeordered products of interacting fields. 
Here we give a local algebraic formulation, i.e. an 
operator identity for a localized interaction. In the case of the variation
of a parameter in the interaction this requires the use of the
retarded product of interacting fields, instead of only time ordered products
(as in the formulation in terms of Green's functions).

For a local construction of observables and physical states in gauge theories
we refer to \cite{DF,BDF}. There, perturbative positivity (``unitarity'')
is, by a local version of the Kugo-Ojima formalism
\cite{KO}, reduced to the validity of BRST symmetry \cite{BRS}. 

\section{Free fields, Borchers' class and local $S$-matrices}

An algebra of observables corresponding to the Klein-Gordon equation
\begin{equation}
(\w +m^2)\varphi =0\label{2.1}
\end{equation}
can be defined as follows: Let $\Delta_{\rm ret,av}$ be the retarded, resp. 
advanced Green's functions of $(\w +m^2)$
\begin{equation}
(\w +m^2)\Delta_{\rm ret,av} =\delta,
\quad\quad {\rm supp}\>\Delta_{\rm ret,av}\subset \bar V_\pm,\label{2.2}
\end{equation}
where $\bar V_\pm$ denotes the closed forward, resp. 
backward lightcone,
and let $\Delta=\Delta_{\rm ret}-\Delta_{\rm av}$. The 
algebra of observables ${\cal A}$ is generated by smeared 
fields $\varphi (f),\>
f\in {\cal D}(\RR^4)$, 
which obey the following relations
\begin{eqnarray}
&&f\mapsto\varphi(f) {\rm \ is \ linear },\label{2.3}\\
&&\varphi((\w +m^2) f)=0,\label{2.4}\\
&&\varphi(f)^{*}=\varphi(\bar f),\label{2.5}\\
&&[\varphi(f),\varphi(g)]=i<f,\Delta *g>.\label{2.6}
\end{eqnarray}
where the star denotes convolution and $<f,g>=\int d^4x 
f(x)g(x)$. As a matter of fact, ${\cal A}$ (as a $*$-algebra 
with unit) is uniquely determined by these relations.

The Fock space representation $\pi$ of the free field is 
induced via the GNS-construction from the vacuum state 
$\omega_0$. Namely, let
$\omega_0:{\cal A}\to \CC$ be the quasifree state given by the 
two-point function
\begin{equation}
\omega_0(\varphi(f)\varphi(g))=i<f,\Delta_+ *g>\label{2.7}
\end{equation}
where $\Delta_+$ is the positive frequency part of $\Delta$. Then the 
Fock space ${\cal H}$, the vector $\Omega$ representing the 
vacuum and the Fock representation are up to equivalence 
determined by the relation
\begin{displaymath}
        (\Omega,\pi(A)\Omega)=\omega_{0}(A)\ ,\ A\in{\cal A}\ .
\end{displaymath}
On ${\cal H}$, the field $\varphi$ (we will omit the representation 
symbol $\pi$) is an operator valued distribution, 
i.e. there is some dense subspace ${\cal D}\subset {\cal H}$ with
\begin{eqnarray}
&&(i)\> \varphi(f)\in {\rm End}({\cal D})\nonumber\\
&&(ii)\> f\mapsto\varphi(f)\Phi\quad\quad {\rm\ is \ continuous}\quad
\forall \Phi\in {\cal D}.\nonumber
\end{eqnarray}
There are other fields $A$ on ${\cal H}$, on the same domain, which
are relatively local to $\varphi$,
\begin{equation}
[A(f),\varphi(g)]=0\quad\quad {\rm if}\quad (x-y)^2<0\quad\forall
(x,y)\in ({\rm supp}\>f\times {\rm supp}\>g).\label{2.8}
\end{equation}
They form the so called Borchers class ${\cal B}$. In the 
case of the free field in 4 dimensions, ${\cal B}$ consists of Wick 
polynomials and their derivatives \cite{E}. Fields from the Borchers 
class can be used
to define local interactions,
\begin{equation}
H_I(t)=-\int d^3x\,g(t,{\vec x})A(t,{\vec x}),\quad\quad g\in 
{\cal D}(\RR^4),\label{2.9}
\end{equation}
(where the minus sign comes from the interpretation of $A$ 
as an interaction term in the Lagrangian)
provided they can be restricted to spacelike surfaces. The corresponding 
time evolution operator from $-\tau$ to $\tau$, 
where $\tau >0$ is so large that ${\rm supp}\>g\,
\subset\,(-\tau ,\tau)\times {\rm R}^3$, (the $S$-matrix) is 
formally given by the Dyson series
\begin{equation}
S(g)={\bf 1}+\sum_{n=1}^\infty\frac{i^n}{n!}\int dx_1...dx_n\,
T\bigl( A(x_1)...A(x_n)\bigr) g(x_1)...g(x_n).\label{2.10}
\end{equation}
with the time ordered products ('$T$-products') $T\bigl(...\bigr)$. 
It is difficult to derive (\ref{2.10}) from (\ref{2.9}) if 
the field $A$ cannot be restricted to spacelike 
surfaces. Unfortunately, this is almost always the case
in four spacetime dimensions, 
the only exception being the field $\varphi$ itself and its 
derivatives. Therefore one defines the timeordered 
products of $n$ factors directly as 
multilinear (with respect to $C^\infty$-functions as coefficients)
symmetric mappings from ${\cal B}^n$ to operator valued distributions
$T\bigl( A_1(x_1)...A_n(x_n)\bigr)$ on ${\cal D}$ such that they 
satisfy the factorization condition\footnote{Due to the symmetry 
and linearity of $T(...)$ it suffices to consider the case
$A_1=A_2=...=A_n$.}
\begin{equation}
T\bigl( A(x_1)...A(x_n)\bigr) =T\bigl( A(x_1)...A(x_k)\bigr)
T\bigl( A(x_{k+1})...A(x_n)\bigr)
\label{2.11}\end{equation}
if $\{x_{k+1},...,x_n\}\,\cap\, (\{x_1,...,x_k\}+\bar V_+)=\emptyset$.
The $S$-matrix $S(g)$ is then, as a formal power 
series,  by definition given by 
(\ref{2.10}) . Since its zeroth order term is ${\bf 1}$, it 
has an inverse in the sense of formal power series 
\begin{equation}
S(g)^{-1}={\bf 1}+\sum_{n=1}^\infty\frac{(-i)^n}{n!}\int dx_1...dx_n\,
\bar T\bigl( A(x_1)...A(x_n)\bigr) g(x_1)...g(x_n),\label{2.12a}
\end{equation}
where the 'antichronological products' $\bar T(...)$ can be 
expressed in terms of the time ordered products
\begin{equation}
\bar T\bigl( A(x_1)...A(x_n)\bigr)\=d\sum_{P\in {\cal P}(\{1,...,n\})}
(-1)^{|P|+n}\prod_{p\in P}T\bigl( A(x_i),i\in p)\ .\label{2.12b}
\end{equation}
Here (${\cal P}(\{1,...,n\})$ is the set of all ordered partitions of 
$\{1,...,n\}$ and $|P|$ is the number of subsets in $P$). The 
$\bar T$-products satisfy anticausal factorization
\begin{equation}
\bar T\bigl( A(x_1)...A(x_n)\bigr) =\bar T\bigl( A(x_{k+1})...
A(x_n)\bigr)\bar T\bigl( A(x_1)...A(x_k)\bigr)
\label{2.12c}\end{equation}
if $\{x_{k+1},...,x_n\}\,\cap\, (\{x_1,...,x_k\}+\bar V_+)=\emptyset$.

The crucial observation now (cf. \cite{IA}) is that $S(g)$
satisfies the remarkable functional equation
\begin{equation}
S(f+g+h)=S(f+g)S(g)^{-1}S(g+h), \label{2.12}
\end{equation}
$f,g,h\in {\cal D}(\RR^4)$,
whenever $({\rm supp}\>f + \bar V_+)\,\cap\,{\rm supp}\>h=\emptyset$
(independent of $g$). Equivalent forms of this equation play an important
role in \cite{BS} and \cite{EG}.
For $g=0$ this is just the functional equation for
the time evolution and may be interpreted as the requirement 
of causality \cite{BS}. Actually, for formal power series 
$S(\cdot )$ of operator valued 
distributions, the $g=0$ equation is equivalent to the seemingly stronger 
relation (\ref{2.12}), because both are equivalent to
condition (\ref{2.11}) for the time
ordered products. We call (\ref{2.12}) the 'condition of causality'.

\section{Interacting local nets}

The arguments of this and the next section are to a large extent 
independent of perturbation theory. 
We start from the assumption that we are given a family 
of unitaries $S(f)\in{\cal A}$, $\forall f\in{\cal D}(\RR^4,{\cal V})$ 
(i.e. $f$ has the form $f=\sum_if_i(x)A_i,\>f_i\in {\cal D}(\RR^4,\RR),
\>A_i\in {\cal V}$) where ${\cal V}$ is an abstract, finite 
dimensional, real vector space, interpreted 
as the space of possible interaction Lagrangians, and ${\cal 
A}$ is some unital $*$-algebra. In perturbation theory ${\cal V}$
is a real subspace of the Borchers' class. The unitaries $S(f)$ are 
required to satisfy the causality condition (\ref{2.12}).
We first observe that we obtain new solutions of (\ref{2.12}) 
by introducing the relative $S$-matrices
\begin{equation} 
S_{g}(f)\=d S(g)^{-1}S(g+f),\label{relS} 
\end{equation}
where now $g$ is kept fixed 
and $S_{g}(f)$ is considered as a functional of $f$. In 
particular, the relative
$S$-matrices satisfy local commutation relations 
\begin{equation}
[S_{g}(h),S_{g}(f)]=0\quad\quad {\rm if}\quad 
(x-y)^2<0\quad\forall (x,y)\in {\rm supp}\>h \times{\rm supp}\>f.\label{3.2}
\end{equation}
Therefore their functional derivatives 
$A_{g}(x)=\frac{\delta}{\delta h(x)}S_{g}(hA)|_{h=0}$, 
$A\in{\cal V}$, $h\in {\cal D}({\RR}^4)$, provided they exist,
are local fields (in the limit $g\rightarrow$ constant
this is Bogoliubov's definition of interactig fields) \cite{BS}.

We now introduce local algebras of observables by 
assigning to
a region ${\cal O}$ of Minkowski space 
the $*$-algebra ${\cal A}_{g}({\cal O})$
which is generated by $\{ S_{g}(h)\>,\>h\in {\cal D}({\cal 
O},{\cal V})\}$.

A remarkable consequence of relation (\ref{2.12}) 
is that the structure of the algebra ${\cal A}_{g}({\cal O})$ 
depends only locally on $g$ \cite{IA,BF}, namely, if
$g\equiv g'$ in a neighbourhood of a causally closed region containing
${\cal O}$, then there exists a unitary $V\in {\cal A}$ such that
\begin{equation}
VS_{g}(h)V^{-1}=S_{g'}(h),\quad\quad\forall \>h\in {\cal 
D}({\cal O},{\cal V}).\label{3.3}
\end{equation}
Hence the system of local algebras of observables (according to the
principles of algebraic quantum field theory this system (``the local
net'') contains the full
physical content of a quantum field theory) is completely determined if one
knows the relative $S$-matrices for test functions 
$g\in {\cal D}(\RR^4,{\cal V})$. 

The construction of the global algebra of
observables  for an interaction Lagrangian ${\cal L}\in{\cal V}$
may be performed explicitly (cf. \cite{BF}). 
Let $\Theta ({\cal O})$
be the set of all functions $\theta\in {\cal D}(\RR^4)$ which
are identically to $1$ in a causally closed open neighbourhood of
${\cal O}$ and consider the bundle
\begin{equation}
\bigcup_{\theta\in \Theta ({\cal O})}\{\theta\}\times {\cal A}_
{\theta {\cal L}}({\cal O}).\label{3.4}
\end{equation}
Let ${\cal U}(\theta,\theta^\prime)$ be the set of all 
unitaries $V\in {\cal A}$ with
\begin{equation}
VS_{\theta{\cal L}}(h)=S_{\theta^\prime {\cal L}}(h)V,\quad\quad\forall 
\>h\in {\cal D}({\cal O},{\cal V}).\label{3.5}
\end{equation}
Then ${\cal A}_{\cal L}({\cal O})$ is defined as the algebra of covariantly 
constant sections, i.e.
\begin{eqnarray}
&&{\cal A}_{\cal L}({\cal O})\ni A=(A_\theta)_{\theta\in \Theta ({\cal O})}
\quad\quad (A_\theta\in {\cal A}_{\theta {\cal L}}({\cal O}))\label{3.6}\\
&&VA_\theta =A_{\theta^\prime}V,\quad\quad\forall V\in  
{\cal U}(\theta,\theta^\prime).\label{3.7}
\end{eqnarray}
${\cal A}_{\cal L}({\cal O})$ contains in particular the elements
$S_{\cal L}(h)$,
\begin{equation}
(S_{\cal L}(h))_\theta =S_{\theta{\cal L}}(h).\label{3.8}
\end{equation}
The construction of the local net is completed by fixing the
embeddings $i_{21}:
{\cal A}_{\cal L}({\cal O}_1)\hookrightarrow {\cal A}_
{\cal L}({\cal O}_2)$ for ${\cal O}_1\subset {\cal O}_2$. But these
embeddings are inherited
from the inclusions ${\cal A}_{\theta {\cal L}}({\cal O}_1)\subset
{\cal A}_{\theta {\cal L}}({\cal O}_2)$ for $\theta\in \Theta ({\cal O}_2)$
by restricting the sections from $\Theta ({\cal O}_1)$ to 
$\Theta ({\cal O}_2)$. The embeddings evidently satisfy the compatibility
relation $i_{12}\circ i_{23}=i_{13}$ for ${\cal O}_3\subset {\cal O}_2
\subset {\cal O}_1$ and define thus an inductive system. Therefore, the 
global algebra can be defined as the inductive limit of local algebras
\begin{equation}
{\cal A}_{\cal L}\=d \cup_{\cal O}{\cal A}_{\cal L}({\cal O}).\label{global}
\end{equation}

In perturbation theory, the unitaries 
$V\in {\cal U}(\theta,\theta^\prime)$ are themselves formal power
series, therefore it makes no sense to say that two elements 
$A,B\in {\cal A}_{\cal L}({\cal O})$ agree {\it in} $n$-th order, 
but only that they agree {\it up to} $n$-th order
(because $(A_\theta-B_\theta)={\cal O}(g^{n+1})$ implies
$A_{\theta^\prime}-B_{\theta^\prime}=
V^{-1}(A_\theta-B_\theta)V={\cal O}(g^{n+1})$).

The time ordered products and hence the relative $S$-matrices 
$S_{\theta {\cal L}}(h)$  are chosen as to 
satisfy Poincar\'e covariance (see the 
normalization condition {\bf N1} below), i.e. the unitary positive
energy representation $U$ of the Poincar\'e group ${\cal
P}_+^\uparrow$ under which the free field transforms satisfies   
\begin{equation} 
U(L)S_{\theta {\cal L}}(h)U(L)^{-1}=S_{\theta_L{\cal L}}(h_L),\quad 
\theta_L(x):=\theta (L^{-1}x),\>h_L(x):=D(L)h(L^{-1}x),\label{3.9} 
\end{equation} 
$\forall L\in {\cal P}_+^\uparrow$ provided ${\cal L}$ is a 
Lorentz scalar and ${\cal V}$ transforms 
under the
finite dimensional representation $D$ of the Lorentz group. 
This enables us to define an automorphic action of
the Poincar\'e group on the algebra of observables. Let for 
$A\in {\cal A}_{\cal L}({\cal O}),\>\theta\in\Theta
(L{\cal O})$ 
\begin{equation}
(\alpha_L(A))_\theta\=d U(L)A_{\theta_{L^{-1}}}U(L)^{-1}.\label{3.10}
\end{equation}
By inserting the definitions one finds that $\alpha_L(A)$ is again a 
covariantly constant section (\ref{3.7}). So $\alpha_L$ is an automorphism
of the net which realizes the Poincar\'e symmetry
\begin{equation} 
\alpha_L{\cal A}_{\cal L}({\cal O})={\cal A}_{\cal L}(L{\cal O}),\quad
\quad  \alpha_{L_1L_2}=\alpha_{L_1}\alpha_{L_2}.
\label{3.10a}\end{equation}

For the purposes of perturbation theory, we have to enlarge 
the local algebras somewhat. In perturbation theory, the 
relative S-matrices are formal power series in two variables, 
and therefore the generators of the local algebras
\begin{equation}
        S_{\cal L}(\lambda f)=\sum_{n=0}^\infty\frac{i^n\lambda^n}{n!}T_{\cal 
        L}(f^{\otimes n})
        \label{E:timeordered products}
\end{equation}
are formal power series with coefficients which are 
covariantly constant sections in the sense of (\ref{3.7}). 
The first order terms in (\ref{E:timeordered products}) are, 
according to Bogoliubov, the interacting local fields,
\begin{equation}
        T_{\cal L}(hA)=:A_{\cal L}(h)\ ,\ A\in{\cal V},\ h\in{\cal 
        D}(\RR^4)\ , 
        \label{E:interacting fields}
\end{equation}
the higher order terms satisfy the causality condition 
(\ref{2.11}) and may therefore be interpreted as time ordered 
products of interacting fields (cf. \cite{EG} sect. 8.1)

Our enlarged local algebra ${\cal A}_{\cal L}({\cal O})$ 
(we use the same symbol as before) now consists of all 
formal power series with coefficients from the algebra 
generated by all timeordered products
$T_{\cal L}(f^{\otimes n})$ with $f\in {\cal D}
({\cal O},{\cal V}),\>n\in\NN_0$. 

\section{Consequences of causality}

Another consequence of the causality relation (\ref{2.12}) 
is that the $S$-matrices $S(f)$ are {\it uniquely} fixed if they 
are known for test functions with arbitrarily small 
supports. Namely, by a repeated use of (\ref{2.12}) we find 
that $S(\sum_{i=1}^n f_{i})$ is a product of factors 
$S(\sum_{i\in K}f_{i})^{\pm 1}$ where the sets $K\subset 
\{1,\ldots,n\}$ have the property that for every pair 
$i,j\in K$ the causal closures of $\supp f_{i}$ and $\supp 
f_{j}$ overlap. Hence if the supports of all $f_{i}$ are 
contained in double cones of diameter $d$, the supports 
of $\sum_{i\in K}f_{i}$ fit into double cones of diameter 
$2d$. As $d>0$ can be chosen arbitrarily small and
the relative $S$-matrices also satisfy (\ref{2.12}), 
this implies  additivity of the net, 
\begin{equation} 
{\cal A}_{\cal L}({\cal O})=\bigvee_\alpha  
{\cal A}_{\cal L}({\cal O}_\alpha)\label{3.11} 
\end{equation} 
where $({\cal O}_\alpha)$ is an arbitrary covering of ${\cal 
O}$ and where the symbol $\bigvee$ means the generated 
algebra.

One might also pose the {\it existence} question: Suppose we have 
a family of unitaries $S(f)$ for all $f$ with sufficiently 
small support which satisfy the causality condition (\ref{2.12}) 
for $f,g,h\in {\cal D}({\cal O},{\cal V})$, diam$({\cal O})$ sufficiently
small, and local commutativity for arbitrary big separation
\begin{displaymath}
        [S(f),S(g)]=0\quad \mbox{if}\quad \supp f\> \mbox{is 
        spacelike to }\supp g\ .
\end{displaymath}
By repeated use of the causality (\ref{2.12}) we can then define 
$S$-matrices for test functions with larger support. It is, 
however, not evident that these $S$-matrices are independent 
of the way of construction and that they satisfy the 
causality condition. (We found a consistent construction only in the
simple case of one dimension: $x=$ time.)
Fortunately, a general positive answer can be 
given in perturbation theory. 
 
Let $S(f)$ be given for $f\in {\cal D}({\cal O},{\cal V})$ 
for all double cones with diam$({\cal O})<r$. The time 
ordered product of $n$ factors is the $n$-fold 
functional derivative of $S$ at $f=0$. It is an operator valued 
distribution\footnote{Here we change the notation for the time
ordered products: let $f=\sum_if_i(x)A_i,\>f_i\in {\cal D}(\RR^4),
\>A_i\in {\cal V}$. Instead of $\int dx_1...dx_n\,\sum_{i_1...i_n}
T\bigl(A_{i_1}(x_1)...A_{i_n}(x_n)\bigr)f_{i_1}(x_1)...f_{i_n}(x_n)$
(\ref{2.10}) we write  $\int dx_1...dx_n\,T_n(x_1,...,x_n)f(x_1)...f(x_n)
\equiv T_n(f^{\otimes n})$.}
$T_{n}$ defined on test functions of $n$ variables with 
support contained in ${\cal U}_n
\=d \{(y_1,...,y_n)\in \RR^{4n}\>|\>{\rm max}_{i<j}|y_i-y_j|<
\frac{r}{2}\}$ and with values in ${\cal V}^{\otimes n}$. 
Especially we know $T_1(x)$ on 
$\RR^4$. On 
this domain the time ordered products satisfy the 
factorization condition (\ref{2.11}). In addition, local 
commutativity of the $S$-matrices implies
\begin{equation}
[T_n(x_1,...,x_n),T_m(y_1,...y_m)]=0\label{2.21}
\end{equation}
for $(x_i-y_j)^2<0\quad\forall (i,j)$ and $(x_1,...x_n)\in {\cal U}_n,\>
(y_1,...,y_m)\in {\cal U}_m$. 
By construction $T_n\vert_{{\cal U}_n}$
is symmetric with respect to permutations of the factors. 

We now show that this input suffices to construct $T_n(x_1,...,x_n)$
on the whole $\RR^{4n}$ by induction on $n$. We assume that the 
$T_k$'s were constructed for $k\leq n-1$, that they fulfil causality 
(\ref{2.11}) and 
\begin{equation}
[T_m(x_1,...,x_m),T_k(y_1,...y_k)]=0\quad {\rm for}\quad (x_1,...x_m)\in 
{\cal U}_m,\>k\leq n-1\label{2.22}
\end{equation}
($m$ arbitrary) and
\begin{equation}
[T_l(x_1,...,x_l),T_k(y_1,...y_k)]=0\quad {\rm for}\quad l,k\leq n-1,
\label{2.23}\end{equation}
if $(x_i-y_j)^2<0\quad\forall (i,j)$ in the latter two equations. 
We can now proceed as in Sect. 4 of \cite{BF}.
\footnote{In contrast 
to the (inductive) Epstein-Glaser construction of $T_n(x_1,...,x_n)$
\cite{EG,BF} the present construction is unique, normalization conditions
(e.g. ${\bf N1}-{\bf N4}$ in sect. 5) are not needed, because the 
non-uniqueness
of the Epstein-Glaser construction is located at the total diagonal 
$\Delta_n \equiv \{(x_1,...,x_n)\> |\> x_1=...=x_n\}$.
But here the time ordered products are given in the neighbourhood 
${\cal U}_n$ of $\Delta_n$.} 

Let ${\cal J}$
denote  the family of all non-empty proper subsets $I$ of the 
index set 
$\{1,...,n\}$ and define the sets ${\cal C}_I\=d \{(x_1,...,x_n)\in
\RR^{4n}\>|\>x_i\not\in J^-(x_j),\,i\in I,\,j\in I^c\}$ for any $I\in 
{\cal J}$. Then
\begin{equation}
\bigcup_{I\in {\cal J}}{\cal C}_I\>\cup\> {\cal U}_n=\RR^{4n}.\label{2.24}
\end{equation}
We use the short hand notations
\begin{equation}
T^I(x_I)=T(\prod_{i\in I}A_i(x_i)),\quad\quad x_I=(x_i,i\in I).\label{2.25}
\end{equation}
On ${\cal D}({\cal C}_I)$ we set
\begin{equation}
T_I(x)\=d T^I(x_I)T^{I^c}(x_{I^c})\label{2.26}
\end{equation}
for any $I\in {\cal C}_I$. For $I_1,I_2\in {\cal J},\>{\cal C}_{I_1}\cap 
C_{I_2}\not=\emptyset$ one easily verifies\footnote{In contrast to \cite{BF}
the Wick expansion of the $T$-products is not used here, because local
commutativity of the $T$-products is contained in the inductive assumption.}
\begin{equation}
T_{I_1}\vert_{{\cal C}_{I_1}\cap {\cal C}_{I_2}}=
T_{I_2}\vert_{{\cal C}_{I_1}\cap {\cal C}_{I_2}}.\label{2.27}
\end{equation}
Let now $\{f_I\}_{I\in {\cal J}}\cup\{f_0\}$ be a finite smooth 
partition of unity of $\RR^{4n}$ subordinate to $\{{\cal C}_I\}_
{I\in {\cal J}}\cup {\cal U}_n$: ${\rm supp}\>f_I\subset {\cal C}_I,
{\rm supp}\>f_0\subset {\cal U}_n$. Then we define
\begin{equation}
T_n(h)\=d T_n\vert_{{\cal U}_n}(f_0h)+\sum_{I\in {\cal J}}T_I(f_Ih),\quad\quad
h\in {\cal D}(\RR^{4n},{\cal V}^{\otimes n}).\label{2.28}
\end{equation}
As in \cite{BF} one may prove that this definition is independent
of the choice of $\{f_I\}_{I\in {\cal J}}\cup\{f_0\}$ and that 
$T_{n}$ is symmetric with respect to 
permutations of the factors and satisfies causality (\ref{2.11}).
Local commutativity (\ref{2.22}) and (\ref{2.23}) (with $n-1$ replaced by $n$)
is verified by inserting the definition (\ref{2.28}) and using
the assumptions. By (\ref{2.10}) we obtain from the $T$-products the 
corresponding $S$-matrix $S(g)$ for arbitrary large support of $g\in
{\cal D}(\RR^4,{\cal V})$, and 
$S(g)$ satisfies the functional equation (\ref{2.12}).

\section{Perturbative quantization and loop expansion}

Causal perturbation theory was traditionally formulated in 
terms of operator valued distributions on Fock space. It is 
therefore well suited for describing the deformation of the 
free field into an interacting field by turning on the 
interaction $g\in{\cal D}(\RR^4,{\cal V})$. It is much less 
clear how an expansion in powers of $\hbar$ can be 
performed, describing the deformation of the classical 
field theory, mainly because the Fock space has no 
classical counter part.
  
Usually the expansion in powers of $\hbar$ is done in functional approaches 
to field theory by ordering Feynman graphs according to 
loop number. In this section we show that the algebraic 
description provides a natural formulation of the loop 
expansion, and we point out the connection to formal quantization theory.  

\subsection{Quantization of a free field and Wick products}

In quantization theory one associates to a given classical theory a 
quantum theory. One procedure is the deformation (or star-product) 
quantization \cite{BFFLS}. This procedure starts from a Poisson algebra,
i.e. a commutative and associative algebra together with a 
second product: a Poisson bracket, satisfying the Leibniz rule and the
Jacobi identity; and to deform the product as a function of $\hbar$, such 
that\footnote{The deformed product is called a $*$-product in deformation
theory. In order to avoid confusion with the $*$-operation we denote 
the product by $\times_\hbar$.} $a\>\times_\hbar\> b$ is a formal power 
series in $\hbar$, the associativity is maintained and
\begin{equation}
a\>\times_\hbar\> b\quad\buildrel\hbar\rightarrow 0\over\longrightarrow\quad 
ab,\quad\quad\quad\frac{1}{\hbar}(a\>\times_\hbar\> b\>-\> b\>\times_\hbar\> a)
\quad\buildrel\hbar\rightarrow 0\over\longrightarrow\quad\{a,b\}.\label{C1}
\end{equation}

Actually this scheme can easily be realized in free field theory
(cf. \cite{Di}). Basic
functions are the evaluation functionals $\varphi_{\rm class} (x)$,
$(\w+m^2)\varphi_{\rm class}=0$, with the Poisson bracket
\begin{equation}
\{\varphi_{\rm class} (x),\varphi_{\rm class} (y)\}=\Delta(x-y)\label{C2}
\end{equation}
($\Delta$ is the commutator function (\ref{2.2})). Because of the singular 
character of $\Delta$ the fields should be smoothed out in order to belong 
to the Poisson algebra. Hence our fundamental classical observables are
\begin{equation}
\phi (t)=t_0+\sum_{n=1}^N\int\varphi_{\rm class}(x_1)...\varphi_{\rm class}
(x_n)t_n(x_1,...,x_n)dx_1...dx_n,\quad t\equiv (t_0,t_1,...),\label{C3}
\end{equation}
where $t_0\in\CC$ arbitrary, $N<\infty$, $t_n$ is a suitable test ``function''
(we will admit also certain distributions) with compact support. 
The Klein Gordon 
equation shows up in the property: $A(t)=0$
if $t_0=0$ and $t_n=(\w_i+m^2)g_n$ for all $n>0$, some $i=i(n)$ and some 
$g_n$ with compact support.

In the quantization procedure we identify  $\varphi_{\rm class}(x_1)...
\varphi_{\rm class}(x_n)$ with the normally ordered product (Wick product) 
$:\varphi(x_1)...\varphi(x_n):$ ($\varphi$ is the free quantum field
(\ref{2.3}-\ref{2.6})). 
Wick's theorem may be interpreted as the definition of a 
$\hbar$-dependent associative product,
\begin{eqnarray}
:\prod_{i\in I}\varphi(x_i):\times_\hbar \>:\prod_{j\in J}\varphi(x_j): &=& 
\nonumber\\
\sum_{K\subset I}\sum_{\alpha:K\rightarrow J\>{\rm injective}}\prod_{j\in K}
i\hbar\Delta_+(x_j-x_{\alpha(j)})&&:\prod_{l\in (I\setminus K)\cup 
(J\setminus\alpha(K))}\varphi(x_l):\label{W2}
\end{eqnarray}
in the linear space spanned by Wick products (the ``Wick 
quantization'').\footnote{The observation that the Wick quantization
is appropriate for the quantization of the free field
goes back to Dito \cite{Di}.} 
To be precise we have to fix a suitable test function 
space (or better: test distribution space) in (\ref{C3}) which is small 
enough such that the product is well defined 
for all $\hbar$ and which contains the interesting cases
occuring in perturbation theory, e.g. products of 
translation invariant distributions (particularly $\delta$-distributions
of difference variables)
with test functions of compact support should be allowed for $t_n$
as in Theorem 0 of Epstein and Glaser. 

Let 
\begin{equation}
{\cal W}_{n}\=d\{t\in {\cal D}^{\prime}(\RR^{4n})_{\rm symm}\>,\>\supp t 
\mbox{ compact },\> 
{\rm WF}(t)\>\cap\>(\RR^{4n}\times\overline{V_+^n\cup V_-^n})=\emptyset\}
\end{equation} 
(see the Appendix for a definition of the wave front set ${\rm WF}$ of 
a distribution). In \cite{BF} it was shown that Wick 
polynomials smeared with distributions $t\in {\cal W}_{n}$, 
\begin{equation}
(\varphi^{\otimes n})(t) \=d\int:\varphi(x_1)...\varphi(x_n):
t(x_1,...,x_n)\,dx_1...dx_n,\quad (\varphi^{\otimes 0})\=d {\bf 1},\label{W3}
\end{equation}
are densely defined operators on an invariant domain in 
Fock space. This includes in particular the Wick powers
\begin{equation}
        :\varphi^n(f):=(\varphi^{\otimes n})(t)\ ,\ f\in{\cal 
        D}(\RR^4)\ ,\ t(x_{1},\ldots,x_{n})=f(x_{1})\prod_{i=2}^n 
        \delta(x_{i}-x_{1})
        \label{E:Wick powers}
\end{equation}
The product of two such operators is given by 
\begin{equation}
(\varphi^{\otimes n})(t)\times_\hbar (\varphi^{\otimes m})(s)=
\sum_{k=0}^{{\rm min}\{n,m\}}\hbar^k(\varphi^{\otimes (n+m-2k)}) 
(t\otimes_{k} s)\label{W6}
\end{equation}
with the $k$-times contracted tensor product
\begin{eqnarray}
(t\otimes_{k} s)(x_1,...,x_{n+m-2k})={\cal S}\frac{n!m!i^k}
{k!(n-k)!(m-k)!}\int dy_1...dy_{2k}\Delta_+(y_1-y_2)...\nonumber\\
\Delta_+(y_{2k-1}-y_{2k})t(x_1,...,x_{n-k},y_1,y_3,...,y_{2k-1})\nonumber\\
s(x_{n-k+1},...,x_{n+m-2k},y_2,y_4,...,y_{2k})\label{W7}
\end{eqnarray}
(${\cal S}$ means the symmetrization in $x_1,...,x_{n+m-2k}$). 
The conditions on the wave front sets of $t$ and 
$s$ imply that the product $(t\otimes_{k} s)$ exists (see the Appendix)
and is an element of ${\cal W}_{n+m-2k}$. The $*$-operation 
reduces to complex conjugation of the smearing function. 

Let ${\cal W}_0\=d\CC$ and ${\cal W}\=d\bigoplus_{n=0}^\infty {\cal W}_{n}$.  
For $t\in {\cal W}$ let $t_n$ denote the component of $t$ in ${\cal W}_{n}$.
The $*$-operation is defined by $(t^*)_{n}\=d(\bar t_{n})$.
Equation (\ref{W6}) can be thought of as the definition of an 
associative product on ${\cal W}$,
\begin{equation}
        (t\times_{\hbar}s)_{n}=\sum_{m+l-2k=n}\hbar^k t_{m}\otimes_{k}s_{l}.
        \label{product on W}
\end{equation}
The Klein-Gordon equation defines an ideal ${\cal N}$ in ${\cal W}$
which is generated by $(\w+m^2)f,\>f\in {\cal D}(\RR^4)$. Actually
this ideal is independent of $\hbar$ (because a contraction with
$(\w+m^2)f$ vanishes) and coincides with the kernel of $\phi$ defined 
in (\ref{C3}). Hence the product (\ref{product on W}) is well defined on 
the quotient space $\bar {\cal W}={\cal W}/{\cal N}$. For a given
positive value of $\hbar$, $\bar {\cal W}$ is isomorphic to the algebra
generated by Wick products $(\varphi^{\otimes n})(t),\>t\in {\cal W}_n$
(\ref{W3}). In the limit $\hbar\rightarrow 0$ we find
\begin{eqnarray}
\lim_{\hbar\to 0}\phi(t)\times_\hbar\phi(s)&=&
\lim_{\hbar\to 0}\phi(\sum_n\hbar^n t\otimes_n s)\nonumber\\
&=& \phi (t\otimes_0 s)=\phi (t)\cdot\phi (s)
\end{eqnarray}
(we set $(t\otimes_k s)_n\=d\sum_{m+l=n}t_{m+k}\otimes_k s_{l+k}$, cf.
(\ref{W7})), with the classical product $\cdot$, and
\begin{equation}
\lim_{\hbar\to 0}\frac{1}{i\hbar}[\phi(t),\phi(s)]_\hbar =
\phi(t\otimes_1 s-s\otimes_1 t)=\{\phi(t),\phi(s)\}
\end{equation}
with the classical Poisson bracket.
Thus $(\bar {\cal W},\times_\hbar)$ provides
a quantization of the given Poisson algebra of the classical free field 
$\varphi_{\rm class}$ (\ref{C2}). We point out that we have formulated
the algebraic structure of smeared Wick products without using the Fock space.
  
The Fock representation is recovered, via the GNS 
construction, from the vacuum state $\omega_0(t)=t_{0}$. It is 
faithful for $\hbar\ne 0$ but is one dimensional in the 
classical limit $\hbar=0$. This illustrates the superiority 
of the algebraic point of view for a discussion of the 
classical limit.

\subsection{Normalization conditions and retarded products}

To study the perturbative quantization of interacting fields we
need some technical tools which are given in this subsection.

The time ordered products are constructed by induction on the number $n$ 
of factors (which is also the order of the perturbation series (\ref{2.10})).
In contrast to the inductive construction of the $T$-products
in sect. 4, we do not know $T_n\vert_{{\cal U}_n}$ here. So
causality (\ref{2.11}) and symmetry determine the time ordered products 
uniquely (in terms of time ordered products of less factors) up to the 
total diagonal $\Delta_n=\{(x_1,...,x_n)\in \RR^{4n}|x_1=x_2=...=x_n\}$.
There is some freedom in the extension to $\Delta_n$. To restrict it we 
introduce the following additional defining
conditions (`normalization conditions',
formulated for a scalar field without derivative
coupling, i.e. ${\cal L}$ is a Wick polynomial solely in $\phi$, it does 
not contain derivatives of $\phi$; for the generalization to
derivative couplings see \cite{BDF})

{\bf N1} covariance with resp. to Poincar\'e transformations and
possibly discrete symmetries, in particular

{\bf N2} unitarity: $T(A_1(x_1)...A_n(x_n))^*=\bar T(A_1^*(x_1)...
A_n^*(x_n))$,

{\bf N3} $\quad [T(A_1(x_1)...A_n(x_n)),\phi (x)]=$

$\quad\quad\quad\quad =i\hbar\sum_{k=1}^n T(A_1(x_1)...
\frac{\d A_k}{\d\phi}(x_k)...A_n(x_n))\Delta (x_k-x)$,

{\bf N4} $\quad (\w_x+m^2)T(A_1(x_1)...A_n(x_n)\phi (x))=$

$\quad\quad\quad\quad\quad\quad =-i\hbar\sum_{k=1}^nT(A_1(x_1)...
\frac{\d A_k}{\d\phi}(x_k)...A_n(x_n))\delta (x_k-x)$
\quad\quad\quad\quad\quad\break
where $[\phi (x),\phi(y)]=i\hbar\Delta (x-y)$. {\bf N1} implies covariance of 
the arising theory, and {\bf N2} provides a $*$-structure. {\bf N3} gives 
the relation to time ordered products of sub Wick polynomials. Once these 
are known (in an inductive procedure), only a scalar distribution has to
be fixed. Due to translation invariance the latter depends only on the 
relative coordinates. Hence, the extension of the 
(operator valued) $T$-product to $\Delta_n$ is reduced to the extension of a 
C-number distribution $t_0\in {\cal D}'(\RR^{4(n-1)}\setminus \{0\})$ to
$t\in {\cal D}'(\RR^{4(n-1)})$. (We call $t$ an extension of $t_0$ if
$t(f)=t_0(f),\>\forall f\in {\cal D}(\RR^{4(n-1)}\setminus \{0\})$). 
The singularity of $t_0(y)$ and $t(y)$
at $y=0$ is classified in terms of Steinmann's scaling degree \cite{SD,BF}
\begin{equation}
{\rm sd}(t)\=d {\rm inf}\{\delta\in \RR\>,\>\lim_{\lambda\to 0}
\lambda^\delta t(\lambda x)=0\}.\label{4.3a}
\end{equation}
By definition ${\rm sd}(t_0)\leq {\rm sd}(t)$, and the 
possible extensions are restricted by requiring
\begin{equation}
{\rm sd}(t_0)={\rm sd}(t).\label{4.3b}
\end{equation}
Then the extension is unique for ${\rm sd}(t_0)<4(n-1)$, and in the 
general case
there remains the freedom to add derivatives of the $\delta$-distribution 
up to order $({\rm sd}(t_0)-4(n-1))$, i.e. 
\begin{equation}
t(y)+\sum_{|a|\leq {\rm sd}(t_0)-4(n-1)}C_a\d^a\delta (y)\label{4.3c}
\end{equation}
is the general solution, where $t$ is a special extension \cite{BF,P,EG},
and the constants $C_a$ are restricted by {\bf N1}, {\bf N2}, {\bf N4}, 
permutation symmetries and possibly further normalization conditions, 
e.g. the Ward identities for QED \cite{DF,BDF}. For an interaction
with mass dimension ${\rm dim}({\cal L})\leq 4$ the requirement (\ref{4.3b})
implies renormalizability by power counting, i.e. the number of indeterminate
constants $C_a$ does not increase by going over to higher perturbative orders.
In \cite{DF} it is shown that the normalization condition {\bf N4} implies 
the field equation for the interacting field corresponding to the free 
field $\phi$ (see also (\ref{4.4}) and sect. 6.1 below).

We have defined the interacting fields as functional derivatives of
relative S-matrices (\ref{E:interacting fields}). Hence, to formulate the perturbation series 
of interacting fields we need the perturbative expansion of the 
relative S-matrices:
\begin{equation}
        S_{g}(f)=\sum_{n,m}\frac{i^{n+m}}{n!m!}R_{n,m}
        (g^{\otimes n};f^{\otimes m}),
        \label{E:retarded products}
\end{equation}
where $g,f\in {\cal D}(\RR^4,{\cal V})$. The 
coefficients are the so called retarded products ('$R$-products'). 
They can be expressed in terms of time ordered and anti-time ordered 
products by
\begin{equation}
        R_{n,m}(g^{\otimes n};f^{\otimes m})=
        \sum_{k=0}^{n}(-1)^k\frac{n!}{k!(n-k)!} 
        \bar{T}_{k}(g^{\otimes k})\times_\hbar
        T_{n-k+m}(g^{\otimes(n-k)} \otimes f^{\otimes m}).\label{R=T}
\end{equation}
They vanish if one of the first $n$ arguments is not in the 
past light cone of some of the last $m$ arguments
(\cite{EG}, sect. 8.1),
\begin{equation}
{\rm supp}\>R_{n,m}\bigl(...\bigr)\subset
\{(y_1,...y_n,x_1,...,x_m)\>,\>\{y_1,...y_n\}\subset (\{x_1,...,x_m\}
+\bar V_-)\} \ .\label{L15a}
\end{equation} 

In the remaining part of this subsect. we show that
the time ordered products can be defined in such a way that 
$R_{n,m}$ is of order $\hbar^n$. For this 
purpose we will introduce the connected part $(a_1\times_\hbar...\times_\hbar
a_n)^c$ of $(a_1\times_\hbar...\times_\hbar a_n)$, where the $a_i$ are 
normally ordered products of free fields, and the connected part
$T_n^c$ of the time ordered product $T_n$
(or 'truncated time ordered product'). In both cases the connected
part corresponds to the sum of connected diagrams, provided the vertices
belonging to the same $a_i$ are identified.
Besides the (deformed) product $\times_\hbar$ 
(\ref{W2}) 
\begin{equation}
a\times_\hbar b=\sum_{n\geq 0}\hbar^n M_n(a,b),\label{W2a}
\end{equation}
where $a,b$ are normally ordered products of free fields,
we have the classical product $a\cdot b=M_0(a,b)$, which is just
the Wick product
\begin{equation}
:\prod_{i\in I}\varphi(x_i):\cdot :\prod_{j\in J}\varphi(x_j): = 
:\prod_{i\in I}\varphi(x_i)\prod_{j\in J}\varphi(x_j):\label{C}
\end{equation}
and which is also associative and in addition commutative. Then we define
$(a_1\times_\hbar...\times_\hbar a_n)^c$ recursively by
\begin{equation}
(a_1\times_\hbar...\times_\hbar a_n)^c\=d (a_1\times_\hbar...
\times_\hbar a_n)
-\sum_{|P|\geq 2}\prod_{J\in P}(a_{j_1}\times_\hbar...\times_\hbar 
a_{j_{|J|}})^c,\label{conn}
\end{equation}
where $\{j_1,...,j_{|J|}\}=J$, $j_1<...<j_{|J|}$,  the sum runs over 
all partitions $P$ of $\{1,...,n\}$ in at least two subsets and
$\prod$ means the classical product (\ref{C}).
$T_n^c$ is defined analogously
\begin{equation}
T_n^c(f_1\otimes...\otimes f_n)\=d T_n(f_1\otimes...\otimes f_n)
-\sum_{|P|\geq 2}\prod_{p\in P}T_{|p|}^c(\otimes_{j\in p}f_j),\label{T^c}
\end{equation}
and similarly we introduce the connected antichronological 
product $\bar T_n^c\equiv (\bar T_n)^c$.

{\bf Proposition 1:} Let the normally ordered products of free fields
$a_1,...,a_n$ be of order ${\cal O} (\hbar^0)$. Then
\begin{equation}
(a_1\times_\hbar...\times_\hbar a_n)^c={\cal O}(\hbar^{n-1}).
\end{equation}

{\bf Proof:} We identify the vertices belonging to the same $a_i$ and
apply Wick's theorem (\ref{W2}) to $a_1\times_\hbar...\times_\hbar a_n$.
Each 'contraction' (i.e. each factor $\Delta_+$) is accompanied by a 
factor $\hbar$. In the terms $\sim\hbar^0$ (i.e. without any contraction)
$a_1,...,a_n$ are completely disconnected, the number of connected 
components is $n$. By a contraction this number is reduced by $1$
or $0$. So to obtain a connected term we need at least $(n-1)$ 
contractions. Hence the connected terms are of order 
${\cal O}(\hbar^{n-1})$. $\quad\w$

Let ${\cal B}\ni A_1,...,A_n={\cal O}(\hbar^0)$ and $x_i\not= x_j,\>
\forall 1\leq i<j\leq n$.
Then there exists a permutation $\pi\in {\cal S}_n$ such that
\begin{equation}
T^c\bigl(A_1(x_1)...A_n(x_n)\bigr)=(A_{\pi 1}(x_{\pi 1})\times_\hbar...
\times_\hbar A_{\pi n}(x_{\pi n}))^c={\cal O}(\hbar^{n-1}).
\label{ordnung:hbar}
\end{equation}
We want this estimate to hold true also for coinciding points
\begin{equation}
T^c\bigl(A_1(x_1)...A_n(x_n)\bigr)={\cal O}(\hbar^{n-1})\quad\quad
{\rm on}\quad {\cal D}(\RR^{4n}).\label{T^c:hbar}
\end{equation}
By the following argument this can indeed be satisfied by appropriate 
normalization of the time ordered products, i.e. (\ref{T^c:hbar})
is an additional normalization condition, which is compatible
with {\bf N1-N4}. We proceed by induction on the number $n$ of factors.
Let us assume that the $T^c$-products with less than $n$ factors fulfil
(\ref{T^c:hbar}) and that we are away from the total diagonal $\Delta_n$.
Using causal factorization, (\ref{T^c}) and the
shorthand notation $T(J):=T(\prod_{j\in J}A_j(x_j)),\> J\subset\{1,...
,n\}$, we then know that there exists $I\subset \{1,...,n\},\>
I\not=\emptyset,\>I^c\not=\emptyset$, with
\begin{eqnarray}
T\bigl(A_1(x_1)...A_n(x_n)\bigr)= T(I)\times_\hbar T(I^c)=
\sum_{r=1}^{|I|}\sum_{s=1}^{|I^c|}\sum_{I_1\sqcup ...\sqcup I_r=I}
\sum_{J_1\sqcup ...\sqcup J_s=I^c}\nonumber\\
\sum_{k\geq 0}\hbar^k
M_k\Bigl(T^c(I_1)\cdot ...\cdot T^c(I_r),
T^c(J_1)\cdot ...\cdot T^c(J_s)\Bigr),\label{fak}
\end{eqnarray}
where $\sqcup$ means the disjoint union. We now pick out the connected 
diagrams. The term $k=0$ on the r.h.s. has $(r+s)$ disconnected 
components. Analogously to Proposition 1 we conclude that it must
hold $k\geq (r+s-1)$ for a connected diagram. Taking the validity of
(\ref{T^c:hbar}) for $T^c(I_l)$ and $T^c(J_m)$ into account, we
obtain $\sum_{l=1}^r (|I_l|-1)+\sum_{m=1}^r (|J_m|-1)+(r+s-1)=n-1$
for the minimal order in $\hbar$ of a connected diagram. So the
$\hbar$-power behaviour (\ref{ordnung:hbar}) holds true on 
${\cal D}(\RR^{4n}\setminus\Delta_n)$, and (\ref{T^c:hbar})
is in fact a normalization condition.

Due to (\ref{T^c}) $(T_n-T_n^c)$ is completely given by timeordered
products of lower orders $<n$ and hence is known also on $\Delta_n$.
The problem of extending $T_n$ to $\Delta_n$ concerns solely $T_n^c$.
The normalization conditions {\bf N1} - {\bf N4} are equivalent to the 
same conditions for $T_n^c$ and $\bar T_n^c$ (i.e. $T_n$ and $\bar T_n$
everywhere replaced by  $T_n^c$ and $\bar T_n^c$). Due to {\bf N3} - 
{\bf N4} it remains
only the extension of $<\Omega,T^c(A_1...A_n)\Omega>$ where all
$A_j$ are different from free fields and $\Omega$ is the vacuum.
It is obvious that this can be done in a way which maintains
(\ref{T^c:hbar}) and is in accordance with {\bf N1} - {\bf N2}.

We emphasize that the (ordinary) time ordered 
product $T_n$ does not satisfy (\ref{T^c:hbar}) because of the presence of
disconneted diagrams. On the other hand the connected antichronological
product $\bar T_n^c$ fulfills the estimate (\ref{T^c:hbar}), as may be
seen by unitarity {\bf N2}. We now turn to the retarded products (\ref{R=T}):

{\bf Proposition 2:} Let ${\cal D}(\RR^4,{\cal V})\ni f_j,g_k=
{\cal O}(\hbar^0)$.
Then the following statements hold true:

(i) All diagrams which contribute to $R_{n,m}(f_1\otimes...\otimes f_n;
g_1\otimes...\otimes g_m)$ have the property that each $f_j$-vertex 
is connected with at least one $g_k$-vertex.

(ii) $R_{n,m}(f_1\otimes...\otimes f_n;g_1\otimes...\otimes g_m)
={\cal O}(\hbar^n)$.

{\bf Proof:} (i) We work with the notation $R_{n,m}(Y;X),\>Y\equiv
\{y_1,...,y_n\},\>X\equiv\{x_1,...,x_m\}$ (cf. \cite{EG}), and consider
a subdiagram with vertices $J\subset Y$ which is not connected with the 
other vertices $(Y\setminus J)\cup X$. Because disconneted diagrams 
factorize with respect to the classical product (\ref{C}), the 
corresponding contribution to $R_{n,m}(Y;X)$ (\ref{R=T}) reads
\begin{equation}
\sum_{I\subset Y}(-1)^{|I|}\Bigl(\bar T(I\cap J^c)\bar T(I\cap J)\Bigr)
\times_\hbar\Bigl(T(I^c\cap J)T(I^c\cap J^c,X)\Bigr).
\end{equation}
However, this expression vanishes due to $\sum_{P\subset J}(-1)^{|P|}
\bar T(P) \times_\hbar T(J\setminus P)=0$ (the latter equation is equivalent 
to (\ref{2.12b}), it is the perturbative version of $S^{-1}S={\bf 1}$).
Hence for non-vanishing diagrams $J$ must be the empty set.

(ii) We express the $R$-product in terms of the connected $T$- and 
$\bar T$-products
\begin{eqnarray}
R_{n,m}(f_1\otimes...\otimes f_n;g_1\otimes...\otimes g_m)=
\sum_{I\subset\{1,...,n\}}(-1)^{|I|}\sum_{P\in {\rm Part}(I)}
\sum_{Q\in {\rm Part}(I^c\sqcup\{1,...,m\})}\nonumber\\
\Bigl(\prod_{p\in P}\bar T_{|p|}^c(\otimes_{i\in p}f_i)\Bigr)\times_\hbar
\Bigl(\prod_{q\in Q}T_{|q|}^c(\otimes_{i\in q}f_i\otimes
\otimes_{j\in q}g_j)\Bigr)
\end{eqnarray} 
where again $\prod$ means the classical product (\ref{C}) and $\sqcup$
stands again for the disjoint union. From (\ref{T^c:hbar}) we know
\begin{equation}
\prod_{p\in P}\bar T_{|p|}^c(\otimes_{i\in p}f_i)={\cal O}(\hbar^{|I|-|P|}),
\quad \prod_{q\in Q}T_{|q|}^c(\otimes_{i\in q}f_i\otimes
\otimes_{j\in q}g_j)={\cal O}(\hbar^{|I^c|+m-|Q|}).\label{prod:hbar}
\end{equation}
From part (i) we conclude that the terms of lowest order (in $\hbar$) in
\begin{equation}
\Bigl(\prod_{p\in P}\bar T_{|p|}^c(...)\Bigr)\times_\hbar
\Bigl(\prod_{q\in Q}T_{|q|}^c(...)\Bigr)=\sum_{n\geq 0}\hbar^n
M_n\Bigl(\prod_{p\in P}\bar T_{|p|}^c(...),\prod_{q\in Q}T_{|q|}^c(...)\Bigr)
\end{equation}
do not contribute. For simplicity we first consider the special case $m=1$.
Then only connected diagrams contribute. Hence we obtain $n\geq |P|+|Q|-1$
similarly to the reasoning after (\ref{fak}). 
For arbitrary $m\geq 1$ the terms 
with minimal power in $\hbar$ correspond to diagrams which are maximally 
disconnected. According to part (i) these diagrams have $m$ disconnected
components each component containing precisely one vertex $g_j$.
Applying the $m=1$-argument to each of this components we get
$n\geq |P|+|Q|-m$. Taking (\ref{prod:hbar}) into account it results the 
assertion: $(|I|-|P|)+(|I^c|+m-|Q|)+(|P|+|Q|-m)=n$. $\quad\w$

\subsection{Interacting fields}

We first describe the perturbative construction of the 
interacting classical field. Let ${\cal L}$ be a function of 
the field which serves as the interaction Lagrangian (for 
simplicity, we do not consider derivative couplings). We 
want to find a Poisson algebra generated by a solution of 
the field equation
\begin{equation}
    (\w+m^2)\varphi_{\cal L}(x)
    =-\Bigl(\frac{\d {\cal L}}{\d\varphi}\Bigr)_{\cal L}(x),
    \label{E:field equation}
\end{equation}
with the initial conditions
\begin{equation}
\begin{split}
        \{\varphi_{\cal L}(0,{\bf x}),\varphi_{\cal L}(0,{\bf y})\}=&0=
        \{\dot{\varphi}_{\cal L}(0,{\bf x}),\dot{\varphi}_{\cal L}
        (0,{\bf y})\}\\
        \{\varphi_{\cal L}(0,{\bf x}),\dot{\varphi}_{\cal L}(0,{\bf y})\}&=
        \delta({\bf x}-{\bf y})\ .
\end{split}     
        \label{E:canonical Poisson brackets}
\end{equation}
We proceed in analogy to the construction of the 
interacting quantum field in Sect. 3 and construct in a 
first step solutions with localized interactions 
$\theta{\cal L}$ with $\theta\in{\cal D}(\RR^4)$ which 
coincide at early times with the free field (hence the initial 
conditions (\ref{E:canonical Poisson brackets}) are trivially satisfied for 
sufficiently early times). They are given 
by a formal power series in the Poisson algebra of the 
free field
\begin{eqnarray}
\varphi_{\theta {\cal L}}(x)=\sum_{n=0}^\infty 
\int_{y_1^0\leq y_2^0
\leq ...y_n^0\leq x^0}dy_1dy_2...dy_n\,\theta (y_1)...\theta (y_n)\nonumber\\
\{{\cal L}(y_1),\{{\cal L}(y_2),...
\{{\cal L}(y_n),\varphi(x)\}...\}\}\label{C4}
\end{eqnarray}
Analogous to the quantum case, the structure of the Poisson algebra 
associated to a causally closed region ${\cal O}$ does not 
depend on the behaviour of the interaction Lagrangian 
outside of ${\cal O}$, i.e. there is, for 
$\theta,\theta'\in \Theta({\cal O})$ a canonical 
transformation $v$ with $v(\varphi_{\theta\cal L}(x))=
\varphi_{\theta'\cal L}(x) $ for all $x\in{\cal O}$. The 
interacting field $\varphi_{\cal L}$ may then be defined as 
a covariantly constant section within a bundle of Poisson algebras. 

Starting from the classical interacting field, one may try 
to define the quantized interacting field by replacing products of 
free classical fields by the normally ordered product of the corresponding
free quantum fields (as in sect. 5.1) and the 
Poisson brackets in (\ref{C4}) by commutators
\begin{equation}
        \{\cdot,\cdot\}\to \frac{1}{i\hbar}[\cdot,\cdot]_{\hbar}\label{coPb}
\end{equation}
where the commutator refers to the quantized product 
$\times_{\hbar}$. Note that in general this replacement produces additional
terms, e.g. the terms $k\geq 2$ in
\begin{eqnarray}
\frac{1}{i\hbar}[:\varphi^n(x):&,&:\varphi^m(y):]_{\hbar}=\sum_{k=1}^{{\rm min}
\>\{n,m\}} (i\hbar)^{k-1}\frac{n!m!}{(n-k)!(m-k)!}\nonumber\\
&&\Bigl(\Delta_+(x-y)^k-\Delta_+(y-x)^k\Bigr)
:\varphi^{(n-k)}(x)\varphi^{(m-k)}(y):
\end{eqnarray}
which correspond to loop diagrams.
Due to the distributional character of 
the fields with respect to the quantized product the 
integral in (\ref{C4}), as it stands, is not well defined
(there is an ambiguity for coinciding points due to the time 
ordering). But as we will see Bogoliubov's formula (\ref{E:interacting fields})
for the interacting quantum field as a functional derivative of 
the relative $S$-matrix 
may be interpreted as a precise version of this integral. 

From the factorization property (\ref{2.11}), (\ref{2.12c}) of time 
ordered and anti-time ordered products, one gets the following 
recursion formula for the retarded products 
(\ref{E:retarded products}-\ref{R=T}):
if $\supp g$ is contained in the past and $\supp f, \supp h$ in the 
future of some Cauchy surface, we find
\begin{equation}
        R_{n+1,m}(g\otimes h^{\otimes n};f^{\otimes m}) = 
        -[T_{1}(g),R_{n,m}(h^{\otimes n};f^{\otimes m})]_\hbar
\label{recursionR}
\end{equation}
where we used the fact that $\bar{T}_{1}=T_{1}$. Hence, for 
$m=1$ and $y_i\not= y_j\>\forall i\not= j$ the retarded product
$R_{n,1}(y_1,...,y_n;x)$ can be written in 
the form\footnote{The notation for the time ordered products introduced 
in section 2 is used here for the retarded products.}
\begin{eqnarray}
R\bigl({\cal L}(y_1)...{\cal L}(y_n);\varphi(x)\bigr)=(-1)^n
\sum_{\pi\in {\cal S}_n}\Theta(x^0-y_{\pi n}^0)\Theta(y_{\pi n}^0-
y_{\pi (n-1)}^0)...\nonumber\\
\Theta(y_{\pi 2}^0-y_{\pi 1}^0)
[{\cal L}(y_{\pi 1}),[{\cal L}(y_{\pi 2})...
[{\cal L}(y_{\pi n}),\varphi(x)]_\hbar...]_\hbar ]_\hbar.\label{R=[,]}
\end{eqnarray}
(Due to the locality of the interaction ${\cal L}$ this is a 
Poincar\'e covariant expression.) This formula confirms
part (ii) of Proposition 2 for non-coinciding $y_i$.
Our main application of (\ref{R=[,]}) is the study of
the classical limit $\hbar\rightarrow 0$ of the quantized 
interacting field (\ref{E:interacting fields}). Due to Proposition 2 
(part (ii)) $R\bigl(\hbar^{-1}{\cal L}(y_1)...\hbar^{-1}{\cal L}
(y_n);\varphi(x)\bigr)$ contains no terms with negative powers of 
$\hbar$ and thus has a well-defined classical limit.
We conclude that the quantized interacting field (\ref{E:interacting fields}), 
(\ref{E:retarded products})
\begin{equation}
\varphi_{\theta{\cal L}}(h)=\sum_{n=0}^\infty\frac{i^n}{n!\hbar^n}
R_{n,1}((\theta{\cal L})^{\otimes n};h\varphi),\quad\quad h\in
{\cal D}(\RR^4),\label{intfield}
\end{equation}
tends to the classical interacting field (\ref{C4}) in this limit.
Note that the factor $\hbar^{-1}$ in the interaction Lagrangian
is in accordance
with the quantization rule (\ref{coPb}), since in (\ref{R=[,]}) there
is for each factor ${\cal L}$ precisely one commutator.
In $R_{n,1}((\theta{\cal L})^{\otimes n};f\varphi)$ the above mentioned 
ambiguities for coinciding points in the iterated retarded 
commutators have been fixed by the definition of time 
ordered products as everywhere defined distributions.

The normalization condition {\bf N4} implies an analogous equation
for the retarded product $R_{n,1}$ (cf. \cite{DF}). The latter means
that $\varphi_{\cal L}$ (\ref{intfield}) satisfies the same
field equation as the classical interacting field (\ref{E:field equation})
\begin {equation}
(\w+m^2)\varphi_{\cal L}(x)=-\Bigl( \frac{\d {\cal L}}{\d\varphi}
\Bigr)_{\cal L}(x).\label{4.4}
\end{equation}
Here $\Bigl( \frac{\d {\cal L}}{\d\varphi}
\Bigr)_{\cal L}$ is not necessarily a polynomial in $\varphi_{\cal L}$ 
(the pointwise product of interacting fields is in general not defined).

We found that the relative $S$-matrices
$S_{\hbar^{-1}\theta{\cal L}}(f)\>(f\in {\cal D}
(\RR^4,{\cal V}))$, and hence all elements of the algebra 
${\cal A}_{\hbar^{-1}\theta\cal L}$ are power series in $\hbar$. 
For the global algebras of covariantly constant sections we recall
from \cite{BF} that the unitaries $V\in {\cal U}(\theta,\theta^\prime)$ 
can be chosen as relative $S$-matrices
\begin{equation} 
V=S_{\hbar^{-1}\theta{\cal L}}(\hbar^{-1}\theta_-{\cal L})^{-1}
\in {\cal U}(\theta,\theta^\prime)\label{E:V=S}
\end{equation}
where $\theta_-\in {\cal D}(\RR^4)$ depends in the following way on 
$(\theta -\theta^\prime)$: we split $\theta -\theta^\prime
=\theta_+ +\theta_-$ with ${\rm supp}\>\theta_+ \cap (C({\cal O})
+\bar V_-)=\emptyset$ and ${\rm supp}\>\theta_- \cap (C({\cal O})
+\bar V_+)=\emptyset$ (where $C({\cal O})$ means the causally closed
region containing ${\cal O}$ in which $\theta$ and $\theta^\prime$ 
agree, cf. (\ref{3.3})).
So $V$ is a formal Laurent series in $\hbar$, and the sections are 
no longer well defined power series. Replacing ${\cal A}$ and ${\cal A}
({\cal O})$ by $\bigvee_{n\in\NN_0}\hbar^n {\cal A}$ and
$\bigvee_{n\in\NN_0}\hbar^n {\cal A}({\cal O})$ (for the new 
algebras the same symbol ${\cal A}$ will be used again) we obtain modules 
over the ring of formal power series in $\hbar$ with complex 
coefficients. For the further construction the validity of part (iii)
of the following Proposition is crucial:

{\bf Proposition 3:}  (i) Let $R_{n,m}(...;...)=\sum_{a=1}^m R_{n,m}^{(a)}
(...;...)$ where $R_{n,m}^{(a)}(...;...)$ is the sum of all diagrams with
$a$ connected components. Then
\begin{equation}
R_{n,m}^{(a)}((\hbar^{-1}\theta{\cal L})^{\otimes n};
(\hbar^{-1}\theta_-{\cal L})^{\otimes m})={\cal O}(\hbar^{-a}).
\label{R^a}
\end{equation}
(Note that the range of $a$ is restricted by part (i) of Proposition 2.)
This estimate is of more general validity: instead of a retarded product 
we could have e.g. a multiple $\times_\hbar$-product, a time ordered or 
antichronological product and the factors may be quite arbitrary. It is only 
essential that each factor is of order ${\cal O}(\hbar^{-1})$.

(ii) Let $A\in {\cal A}({\cal O})$. Then all diagrams which contribute to
$V\times_\hbar A\times_\hbar V^{-1}$ (where $V$ is given by (\ref{E:V=S}))
have the property that each vertex of $V$ and of $V^{-1}$ is 
connected with at least one vertex of $A$. (It may happen that a connected
component of $V$ is not directly connected with $A$, but that it is 
connectecd with a connected component of $V^{-1}$ and the latter is 
connected with $A$.)

(iii)  \begin{equation}
{\cal A}({\cal O})\ni A={\cal O}(\hbar^n)\quad\Longrightarrow
\quad V\times_\hbar A\times_\hbar V^{-1}={\cal O}(\hbar^n)\label{E:Ad(V)A}
\end{equation}
In particular if $A$ is the term of $n$-th order in $\hbar$ of an interacting 
field, then $V\times_\hbar A\times_\hbar V^{-1}$ is a power series in 
$\hbar$ in which the terms up to order $\hbar^{n-1}$ vanish.

{\bf Proof:} Part (i) is obtained essentially in the same way as Proposition 
1. Part (iii) is a consequence of parts (i) and (ii), and the following 
observation: let us consider a diagram which contributes to $V\times_\hbar 
A\times_\hbar V^{-1}$ according to part (ii). If the subdiagrams belonging 
to $V$ and $V^{-1}$ have $r$ and $s$ connected components, then the 
whole diagram has at least $(r+s)$ contractions, which yield a factor 
$\hbar^{(r+s)}$.

It remains the proof of (ii): We use the same notations as in the proof 
of Proposition 2. Let $Y_1\sqcup Y_2=Y$, $X_1\sqcup X_2=X$.
We now consider the sum of all diagrams contributing
to $R(Y,X)$ in which the vertices $(Y_1,X_1)$ are not connected with the 
vertices $(Y_2,X_2)$. Using (\ref{R=T}) and the fact that disconnected
diagrams factorize with respect to the classical product (\ref{C}),
this (partial) sum is equal to
\begin{eqnarray}
\sum_{I\subset Y}(-1)^{|I\cap Y_1|}[\bar T(I\cap Y_1)\times_\hbar
T(I^c\cap Y_1,X_1)]\cdot\nonumber\\
(-1)^{|I\cap Y_2|}[\bar T(I\cap Y_2)\times_\hbar
T(I^c\cap Y_2,X_2)]\nonumber\\
=R(Y_1,X_1)\cdot R(Y_2,X_2).\label{RR}
\end{eqnarray}
From ${\bf 1}=VV^{-1}=VV^*$, (\ref{E:retarded products}) and 
(\ref{E:V=S}) we know
\begin{equation}
\sum_{Y_1\sqcup Y_2=Y,\>X_1\sqcup X_2=X}(-1)^{(|Y_1|+|X_1|)}
R^*(Y_1,X_1)\times_\hbar R(Y_2,X_2)=0\label{V^*V}
\end{equation}
for fixed $(Y,X)$, $Y\cup X\not=\emptyset$. Next we note
\begin{eqnarray}
V\times_\hbar A\times_\hbar V^{-1}=\sum_{n,m}\frac{1}{n!m!}\int
dy_1...dy_ndx_1...dx_m\,\theta(y_1)...\theta(y_n)\theta_-(x_1)...
\theta_-(x_m)\nonumber\\
\sum_{Y_1\sqcup Y_2=Y,\>X_1\sqcup X_2=X}
(-i)^{(|Y_1|+|X_1|)}i^{(|Y_2|+|X_2|)}R^*(Y_1,X_1)\times_\hbar A
\times_\hbar R(Y_2,X_2),\label{VAV^*}
\end{eqnarray}
where we have used the notations $Y\equiv\{y_1,...y_n\},\>
X\equiv\{x_1,...,x_n\}$. In the integrand of the latter expression
we consider (for given $Y$ and $X$) fixed decompositions $Y=Y_3\sqcup Y_4$
and $X=X_3\sqcup X_4$, $Y_3\cup X_3\not=\emptyset$. 
Now we consider the (partial) 
sum of all diagrams in which the vertices $(Y_3,X_3)$ are not connected
with $A$ and each of the vertices $(Y_4,X_4)$ is connected with $A$.
Part (ii) is proved if we can show that this partial sum vanishes. This 
holds in fact true because $R^*$ and $R$ factorize according to 
(\ref{RR}), and due to the unitarity (\ref{V^*V}):
\begin{eqnarray}
\sum_{Y_1\sqcup Y_2=Y,\>X_1\sqcup X_2=X}(-1)^{(|Y_1\cap Y_4|+|X_1\cap X_4|)}
[R^*(Y_1\cap Y_4,X_1\cap X_4)
\times_\hbar \nonumber\\
A\times_\hbar R(Y_2\cap Y_4,X_2\cap X_4)]\cdot\nonumber\\
(-1)^{(|Y_1\cap Y_3|+|X_1\cap X_3|)}
[R^*(Y_1\cap Y_3,X_1\cap X_3)
\times_\hbar R(Y_2\cap Y_3,X_2\cap X_3)]=0.\quad\w\nonumber
\label{VAV^*=0}
\end{eqnarray}

Now we are ready to give an algebraic formulation of the expansion 
in $\hbar$. Let $I_{n}\=d\hbar^{n} {\cal A}_{\cal L}$. 
$I_{n}$ is an ideal in the global algebra ${\cal A}_{\cal L}$. We define
\begin{equation}
{\cal A}_{\cal L}^{(n)}\=d\frac{{\cal A}_{\cal L}}{I_{n+1}},\quad\quad
{\cal A}_{\cal L}^{(n)}({\cal O})\=d\frac{{\cal A}_{\cal L}
({\cal O})}{I_{n+1}\cap {\cal A}_{\cal L}({\cal O})}.\label{L16a}
\end{equation}
which means that we neglect all terms which are of order ${\cal O}
(\hbar^{n+1})$. The embeddings $i_{21}:
{\cal A}_{\cal L}({\cal O}_1)\hookrightarrow {\cal A}_
{\cal L}({\cal O}_2)$ for ${\cal O}_1\subset {\cal O}_2$ 
induce embeddings ${\cal A}_{\cal L}^{(n)}({\cal O}_1)\hookrightarrow 
{\cal A}_{\cal L}^{(n)}({\cal O}_2)$. Thus we obtain a 
projective system of local nets $({\cal A}_{\cal 
L}^{(n)}({\cal O}))$ of algebras of quantum 
observables up to order $\hbar^{n+1}$.

Note that we may equip our algebras ${\cal A}_{\cal L}^{(n)}$
also with the Poisson 
bracket induced by $\frac{1}{i\hbar}[\cdot,\cdot]_{\hbar}$, 
because the ideals $I_{n}$ are also Poisson ideals with 
respect to these brackets. 
Then ${\cal A}_{\cal L}^{(0)}$ becomes the local net of 
Poisson algebras of the classical field theory, whereas for 
$n\ne 0$  we obtain a 
net of noncommutative Poisson algebras.  

The expansion in powers of $\hbar$ is usually called 
``loop expansion''. This is
due to the fact that the order
in $\hbar$ of a certain Feynman diagram belonging to 
$R_{n,m}((\hbar^{-1}\theta{\cal L})^{\otimes n};
f_1\otimes ...\otimes f_m),
\>{\cal D}(\RR^4,{\cal V})\ni f_j={\cal O}(\hbar^0)$, is equal to: 
(number of propagators (i.e. inner lines)) - $n$ $=$
(number of loops) + $m$ - (number of connected components). 
In particular, using part (i) of Proposition 2, we find
that for the interacting fields ($m=1$) the order in $\hbar$
agrees with the number of loops.

\section{Local algebraic formulation of the quantum action principle}

The method of algebraic renormalization (for an overview see 
\cite{PS}) relies on the so called 'quantum action principle' (QAP), 
which is due 
to Lowenstein \cite{L} and Lam \cite{Lam}. This principle is a formula for 
the variation of (possibly connected or one-particle-irreducible) Green's
functions (or of the corresponding generating functional) under

- a change of coordinates (e.g. one applies the differential operator of
the field equation to the Green's functions),

- a variation of the fields (e.g. the BRST-transformation)

- a variation of a parameter. This may be a parameter in the Lagrangian
or in the normalization conditions for the Green's functions.

These are different theorems with different proofs. The common statement
is that the variation of the Green's functions is equal to the insertion
of a local or spacetime integrated composite field operator
(for details see \cite{PS}). In this section we study two simple cases
of the QAP: the field equation and the variation of 
a parameter which appears only in the interaction Lagrangian. 

The aim of this section is to formulate the QAP (in these two cases) for 
our local algebras of observables ${\cal G}_{\cal L}({\cal O})$, i.e.
we are looking for an {\it operator} identity which holds true independently 
of the adiabatic limit. Such an identity does not depend on the choice
of a state, as it is the case for the Green's functions.

In a second step we compare our formula
with the usual formulation of the QAP in terms 
of Green's functions. The latter are the vacuum expectation values
in the adiabatic limit $g\rightarrow 1$.\footnote{This limit is taken by
scaling the test function $g$: let $g_0\in {\cal D}(\RR^4),\>g_0(0)=1$; then
one considers the limit $\epsilon\rightarrow 0\>\>(\epsilon >0)$
of $g_\epsilon(x)\equiv g_0(\epsilon x)$. Uniqueness of the adiabatic limit
means the independence of the particular choice of $g_0$.} We specialize 
to models for which the adiabatic limit is known to exist. This is 
the case for pure massive theories \cite{EG} and certain theories with (some)
massless particles such as QED and $\lambda :\varphi^{2n}:$-theories 
\cite{BlSe}, provided the time ordered products are appropriately normalized.

{\it Remarks}: (1) From the usual QAP (in terms of Green's functions) 
one obtains an operator 
identity by means of the Lehmann-Symanzik-Zimmermann - reduction 
formalism \cite{LSZ}. Although the latter
relies on the adiabatic limit an analogous
conclusion from the Fock vacuum expectation values to arbitrary matrix
elements is possible in our local construction: let ${\cal O}$ be
an open double cone and let $x_1,...,x_k\not\in ((\bar {\cal O}\cup
\{x_{k+l+1},...,x_n\})+\bar V_-),\>x_{k+1},...,x_{k+l}
\in {\cal O}$ and $x_{k+l+1},...,x_n\not\in (\bar {\cal O}+\bar V_+)$. 
Using the causal factorization
of time ordered products of interacting fields (\ref{E:timeordered 
products}) we obtain
\begin{eqnarray}
\Bigl(\Omega ,T_{\theta{\cal L}}\bigl(\varphi(x_1)...\varphi(x_n)\bigr)
\Omega\Bigr)=
\Bigl(T_{\theta{\cal L}}\bigl(\varphi(x_1)...\varphi(x_k)\bigr)^*\Omega ,
\nonumber\\
T_{\theta{\cal L}}\bigl(\varphi(x_{k+1})...\varphi(x_{k+l})\bigr)
T_{\theta{\cal L}}\Bigl(\varphi(x_{k+l+1})...\varphi(x_n)\bigr)\Omega\Bigr).
\end{eqnarray}
Now we choose $\theta\in \Theta({\cal O})$ such that $\{x_1,...,x_k\}\cap
({\rm supp}\>\theta+\bar V_-)=\emptyset$ and $\{x_{k+l+1},...,x_n\}\cap
({\rm supp}\>\theta+\bar V_+)=\emptyset$. Due to the retarded support
(\ref{L15a}) of the $R$-products we then know that
$T_{\theta{\cal L}}\bigl(\varphi(x_{k+l+1})...\varphi(x_n)\bigr)$ agrees 
with the time ordered product $T_0\bigl(\varphi(x_{k+l+1})...
\varphi(x_n)\bigr)$ of the corresponding free fields. By means of
$S_{\theta{\cal L}}(f\varphi)=S(\theta{\cal L})^{-1}S(f\varphi)
S(\theta{\cal L})$
for ${\rm supp}\>f\cap ({\rm supp}\>\theta+\bar V_-)=\emptyset$ we obtain
\begin{equation}
T_{\theta{\cal L}}\bigl(\varphi(x_1)...\varphi(x_k)\bigr)^*=S(\theta
{\cal L})^{-1}T_0\bigl(\varphi(x_1)...\varphi(x_k)\bigr)^*
S(\theta{\cal L}).
\end{equation}
Our assertion follows now from the fact that the states $T_0\bigl(
\varphi(x_{k+l+1})...\varphi(x_n)\bigr)\Omega$  generate a dense subspace 
of the Fock space and the same for the states \break $S(\theta{\cal L})^{-1}T_0
\bigl(\varphi(x_1)...\varphi(x_k)\bigr)^*S(\theta{\cal L})\Omega$. (For 
the validity of the latter statement it is important that $x_1,...,x_k$
can be arbitrarily spread over a Cauchy surface which is later
than $(\bar {\cal O}\cup\{x_{k+l+1},...,x_n\})$.)

(2) Recently Pinter \cite{Pi} presented an alternative 
derivation of
the QAP for the variation of a parameter in the Lagrangian (including the 
free part) also in the framework of causal perturbation theory. 
In contrast to our
presentation Pinter's QAP is formulated for the $S$-matrix and in the 
adiabatic limit. The main new technical tool which is used is a 
generalization of the normalization condition {\bf N4}.

\subsection{Field equation}

The normalization condition {\bf N4} implies
\begin {eqnarray}
(\w_x+m^2)R\bigl({\cal L}(y_1)...{\cal L}(y_n);\phi(x)\phi(x_1)...
\phi(x_m)\bigr)=\nonumber\\
-i\sum_{l=1}^n\delta(x-y_l)R\bigl({\cal L}(y_1)...\hat l...{\cal L}(y_n);
\frac{\d {\cal L}}{\d\phi}(x)\phi(x_1)...\phi(x_m)\bigr)\nonumber\\
-i\sum_{j=1}^m\delta(x-x_j)R\bigl({\cal L}(y_1)...{\cal L}(y_n);\phi(x_1)...
\hat j...\phi(x_m)\bigr),\label{4.5}
\end{eqnarray}
where $\hat l$ and $\hat j$ means that the corresponding factor is omitted.
This equation takes a simple form for the corresponding generating 
functionals (i.e. the relative $S$-matrices (\ref{relS}))
\begin {equation}
f(x)S_{g{\cal L}}(f\phi)=(\w_x+m^2)\frac{\delta}{i\delta f(x)}
S_{g{\cal L}}(f\phi)
-\frac{\delta}{i\delta \rho(x)}\vert_{\rho =0}S_{g{\cal L}}(f\phi
+\rho g\frac{\d {\cal L}}{\d\phi}).\label{4.5a}
\end{equation}
To formulate this in terms of our local algebras of observables 
(cf. sect. 3) we set $g\equiv\theta\in \Theta({\cal O})$ and for
$x\in {\cal O}$ we can choose $\rho$ such that ${\rm supp}\>\rho
\subset\{y|\theta (y)=1\}$. Then (\ref{4.5a}) turns into
\begin{equation}
(\w_x+m^2)\frac{\delta}{i\delta f(x)}
S_{\cal L}(f\phi)=f(x)S_{\cal L}(f\phi)+
\frac{\delta}{i\delta \rho(x)}\vert_{\rho =0}S_{\cal L}(f\phi
+\rho \frac{\d {\cal L}}{\d\phi}),\quad x\in {\cal O}.\label{4.5aa}
\end{equation}
This is the QAP (in the case of the field equation) for the local algebras 
of observables.

To compare with the usual form of the QAP we consider
the generating functional $Z(f)$ for the Green's functions $<\Omega|
T\bigl(\phi_{\cal L}(x_1)...\phi_{\cal L}(x_m)\bigr)|\Omega >$ which is 
obtained from the relative $S$-matrices by
\begin{equation}
Z(f)=\lim_{g\to 1}(\Omega,S_{g{\cal L}}(f\phi)\Omega),\label{4.5b}
\end{equation}
where $\Omega$ is the Fock vacuum \cite{EG}. So by taking
the vacuum expectation value and the adiabatic limit of (\ref{4.5a})
we get
\begin{equation}
f(x)Z(f)=-\Delta (x)\cdot Z(f),\label{4.5c}
\end{equation}
where $\Delta (x)$ is a insertion of UV-dimension\footnote{We assume that
${\cal L}$ has UV-dimension $4$.} $3$, coinciding with the 
classical field polynomial $\frac{\delta S}{\delta\phi (x)}$ in the 
classical approximation (where $S=\int d^4x\,
[\frac{1}{2}(\d_\mu\phi(x)\d^\mu\phi(x)-m^2\phi^2(x))+g(x){\cal L}(x)]$ 
is the classical action). Equation (\ref{4.5c}) is the usual form of the
QAP (cf. eqn. (3.20) in \cite{PS}). In the present case the local 
algebraic formulation (\ref{4.5aa}) contains more information than the 
usual QAP (\ref{4.5c}).

\subsection{Variation of a parameter in the interaction}

In (\ref{E:retarded products}) we have defined retarded products of Wick
polynomials, i.e. elements of the Borchers class. Analogously we now
introduce retarded products $R_{\cal L}(g^{\otimes n};f^{\otimes m})$
of interacting fields
\begin{equation}
        S_{{\cal L}+g}(f)=S_{\cal L}(g)^{-1}S_{\cal L}(g+f)\=d
\sum_{n,m=0}^\infty\frac{i^{n+m}}{n!m!}R_{\cal L}
        (g^{\otimes n};f^{\otimes m})
        \label{Q2b}
\end{equation}
where ${\cal L},g,f\in {\cal D}(\RR^4,{\cal V})$. Obviously they can be
expressed in terms of antichronological and time ordered products
of interacting fields by exactly the same formula as in the case of Wick
polynomials (\ref{R=T})
\begin{equation}
        R_{\cal L}(g^{\otimes n};f^{\otimes m})=
        \sum_{k=0}^{n}(-1)^k\frac{n!}{k!(n-k)!} 
        \bar{T}_{\cal L}(g^{\otimes k})
        T_{\cal L}(g^{\otimes(n-k)} \otimes f^{\otimes m}).\label{R=Tww}
\end{equation}
Thereby the antichronological product of interacting fields is defined
analogously to the time ordered product (\ref{E:timeordered products}), 
namely by
\begin{equation}
\bar T_{\cal L}(f^{\otimes m})=
\frac{d^m}{(-i)^m d\lambda^m}\vert_{\lambda =0}S_{\cal L}
(\lambda f)^{-1},\label{Q2a}
\end{equation}
and satisfies anticausal factorization (\ref{2.12c}) (which 
justifies the name). The support property (\ref{L15a}) of the
retarded products relies on the 
(anti)causal factorization of the $T$- and $\bar T$-products (\ref{2.11},
\ref{2.12c}), hence, the 
$R$-product of interacting fields (\ref{Q2b}-\ref{R=Tww}) has also retarded 
support (\ref{L15a}).

Similarly to Lowenstein in \cite{L} sect.II.B we consider an infinitesimal 
change of the interaction Lagrangian
\begin{equation}
{\cal L}_0\rightarrow {\cal L}_0+\epsilon{\cal L}_1\label{Q1}
\end{equation}
where ${\cal L}_0,{\cal L}_1\in {\cal V}$ or ${\cal D}(\RR^4,{\cal V})$.
For the $m$-fold variation of the time ordered product of the interacting 
fields (\ref{E:timeordered products}) we obtain
\begin{eqnarray}
\frac{d^m}{d\epsilon^m}\vert_{\epsilon =0}
T_{\theta ({\cal L}_0+\epsilon {\cal L}_1)}(f^{\otimes l})&=&
\frac{\d^m}{\d\epsilon^m}\vert_{\epsilon =0}\frac{\d^l}{i^l\d\lambda^l}
\vert_{\lambda =0}S_{\theta ({\cal L}_0+\epsilon {\cal L}_1)}(\lambda f)
\nonumber\\
&=&i^m R_{\theta {\cal L}_0}((\theta {\cal L}_1)^{\otimes m};f^{\otimes l}).
\label{Q2}
\end{eqnarray}

To formulate this identity for our local algebras 
of observables we assume that the ${\cal L}_1$ has compact 
support, i.e. ${\cal L}_1\in {\cal D}(\RR^4,{\cal V})$. We set
\begin{equation}
\Theta_0({\cal O})\=d\{\theta\in \Theta({\cal O})\quad |\quad
\theta\vert_{{\rm supp}\> {\cal L}_1}\equiv 1\}.
\end{equation}
We consider the observables as covariantly constant sections in the bundle
over $\Theta_0({\cal O})$ (instead of $\Theta ({\cal O})$ as in sect. 3).
Then we obtain
\begin{equation}
\frac{d^m}{d\epsilon^m}\vert_{\epsilon =0}
T_{{\cal L}_0+\epsilon {\cal L}_1}(f^{\otimes l})=
i^m R_{{\cal L}_0}({\cal L}_1^{\otimes m};f^{\otimes l}).\label{Q2c}
\end{equation}
This is the local algebraic 
formulation of the QAP for the variation of a parameter in the interaction.

We are now going to investigate the usual QAP by using Epstein and Glaser's
definition of Green's functions (\ref{4.5b}).
In (\ref{Q2}) the $m$-fold variation of the parameter $\epsilon$ results in 
a {\it retarded} insertion of $(\theta{\cal L}_1)^{\otimes m}$. In the usual
QAP $(\theta{\cal L}_1)^{\otimes m}$ is inserted into the {\it time ordered}
product, i.e. one considers
\begin{equation}
i^m T_{\theta {\cal L}_0}((\theta {\cal L}_1)^{\otimes m}
\otimes f^{\otimes l})=\frac{\d^m}{\d\epsilon^m}\vert_{\epsilon =0}
\frac{\d^l}{i^l\d\lambda^l}\vert_{\lambda =0}S_{\theta {\cal L}_0}
(\theta\epsilon {\cal L}_1+\lambda f).\label{Q3}
\end{equation}
Obviously (\ref{Q2}) and (\ref{Q3}) do not agree. However, let us assume 
that we are dealing with a pure massive theory and that ${\cal L}_0$ and
${\cal L}_1$ have UV-dimension ${\rm dim}({\cal L}_j)=4$. Or: if 
${\rm dim}({\cal L}_j)<4$ we assume that ${\cal L}_j$ is treated in the
extension to the total diagonal as if it would hold ${\rm dim}({\cal L}_j)
=4$. Hence it may occur that the scaling degree increases in the extension
to a certain amount: ${\rm sd}(t_0)\leq {\rm sd}(t)\leq 4n-b$
for a scalar theory without derivative couplings, where $b$ is the number
of external legs (cf. (\ref{4.3a})-(\ref{4.3c})). (In the BPHZ framework 
one says that ${\cal L}_j$ is 'oversubtracted with degree $4$'.)
Then there exists a normalization of the time ordered products,
which is compatible with the other normalization conditions {\bf N1} -
{\bf N4} and (\ref{T^c:hbar}), such that the Green's functions 
corresponding to (\ref{Q3}) exist and agree, i.e. we assert
\begin{equation}
\frac{d^m}{d\epsilon^m}\vert_{\epsilon =0}\lim_{\theta\to 1}
\Bigl(\Omega,T_{\theta ({\cal L}_0+\epsilon {\cal L}_1)}(f^{\otimes l})
\Omega\Bigr)=i^m\lim_{\theta\to 1}\Bigl(\Omega,T_{\theta {\cal L}_0}
((\theta {\cal L}_1)^{\otimes m}\otimes f^{\otimes l})\Omega\Bigr)\label{Q5}
\end{equation}
for all $m,l\in \NN_0$, which is equivalent to
\begin{equation}
\lim_{\theta\to 1}\Bigl(\Omega,S_{\theta ({\cal L}_0+\epsilon {\cal L}_1)}
(\lambda f)\Omega\Bigr)=\lim_{\theta\to 1}\Bigl(\Omega,S_{\theta {\cal L}_0}
(\theta\epsilon {\cal L}_1+\lambda f)\Omega\Bigr).\label{Q6}
\end{equation}
(We assume that the derivatives $\frac{\d^m}{\d\epsilon^m}$ and
$\frac{\d^l}{\d\lambda^l}$ commute with the adiabatic limit
$\theta\rightarrow 1$. This seems to be satisfied for vacuum expectation 
values in pure massive theories as it is the case here \cite{EG}.) 
This is the usual form of the QAP (in terms of Epstein and Glaser's
Green's functions) for the present case (cf. eqn. (2.6)
of \cite{L} \footnote{Lowenstein works with Zimermanns definition of normal 
products of interacting fields: $N_\delta \{\prod_{j=1}^l\varphi_{i_j\>
{\cal L}}(x)\},\>\delta\geq d\equiv \sum_{j=1}^l d(\varphi_{i_j\>{\cal L}})$ 
\cite{Z}. For $\delta =d$ (i.e. without oversubtraction) $N_\delta 
\{\prod_{j=1}^l\varphi_{i_j\>{\cal L}}(x)\}$ agrees essentially with our
$(:\prod_{j=1}^l\varphi_{i_j}(x):)_{g{\cal L}}$ (\ref{E:interacting fields}). 
The difference
is due to the adiabatic limit and the different ways of defining Green's 
functions (Zimmermann uses the Gell-Mann Low series, cf. (\ref{Q4}),
(\ref{gml4})).}). In contrast to the field equation, the QAP 
(\ref{Q5}) does not hold for the operators before the adiabatic limit.

{\it Proof of (\ref{Q5})}: For a better comparison with Lowenstein's
formulation, we present a proof which makes the detour over the 
corresponding Gell-Mann Low expressions. First we comment 
on the equality of Epstein and Glaser's
Green's functions with the Gell-Mann Low series
\begin{equation}
\lim_{\theta\to 1}(\Omega,S_{\theta{\cal L}}(f)\Omega)=
\lim_{\theta\to 1}\frac{(\Omega,S(\theta{\cal L}+f)\Omega)}
{(\Omega,S(\theta{\cal L})\Omega)},\label{Q4}
\end{equation}
which is proved in the appendix of \cite{D}. This can be understood in 
the following way: let $P_\Omega$ be the projector on the Fock vacuum
$\Omega$ and $P_\Omega^\bot\=d 1-P_\Omega$. Using $S(\theta{\cal L})^*
=S(\theta{\cal L})^{-1}$ we obtain
\begin{eqnarray}
(\Omega,S_{\theta{\cal L}}(f)\Omega)&=&(S(\theta{\cal L})\Omega,
(P_\Omega+P_\Omega^\bot)S(\theta{\cal L}+f)\Omega)\nonumber\\
&=&\frac{(\Omega,S(\theta{\cal L}+f)\Omega)}{(\Omega,S(\theta{\cal L})\Omega)}
\cdot |(\Omega,S(\theta{\cal L})\Omega)|^2\nonumber\\
&+&(\Omega,S(\theta{\cal L})^{-1}
P_\Omega^\bot S(\theta{\cal L}+f)\Omega)\label{gml1}
\end{eqnarray}
and
\begin{eqnarray}
1&=&(\Omega,S(\theta{\cal L})^{-1}(P_\Omega+P_\Omega^\bot)S(\theta{\cal L})
\Omega)\nonumber\\
&=&|(\Omega,S(\theta{\cal L})\Omega)|^2+
(\Omega,S(\theta{\cal L})^{-1}P_\Omega^\bot S(\theta{\cal L})\Omega).
\label{gml2}\end{eqnarray}
In $(\Omega,S(\theta{\cal L})^{-1}P_\Omega^\bot S(\theta{\cal L}+f)\Omega)$
there is at least one contraction between $S(\theta{\cal L})^{-1}$ and
$S(\theta{\cal L}+f)$ (or: the terms without contraction are precisely
$(\Omega,S(\theta{\cal L})^{-1}\Omega)\break
(\Omega, S(\theta{\cal L}+f)\Omega)$).
In the mentioned reference the support properties in momentum space of
the contracted terms are analysed and in this way it is proved
\begin{equation}
\lim_{\theta\to 1}(\Omega,S(\theta{\cal L})^{-1}P_\Omega^\bot 
S(\theta{\cal L}+f)\Omega)=0.\label{gml3}
\end{equation}
Inserting this into (\ref{gml1}) and (with $f=0$) into (\ref{gml2})
it results (\ref{Q4}).

Because of (\ref{Q4}) our assertion (\ref{Q6}) is equivalent to
\begin{equation}
\lim_{\theta\to 1}\frac{(\Omega,S(\theta({\cal L}_0+\epsilon {\cal L}_1)
+\lambda f)\Omega)}
{(\Omega,S(\theta ({\cal L}_0+\epsilon {\cal L}_1))\Omega)}
=\lim_{\theta\to 1}\frac{(\Omega,S(\theta({\cal L}_0+\epsilon {\cal L}_1)
+\lambda f)\Omega)}
{(\Omega,S(\theta {\cal L}_0)\Omega)}.\label{gml4}
\end{equation}
This is the QAP in terms of the Gell-Mann Low series. Obviously
the nontrivial statement is
\begin{equation}
\lim_{\theta\to 1}\frac{(\Omega,S(\theta({\cal L}_0+\epsilon {\cal L}_1))
\Omega)}{(\Omega,S(\theta {\cal L}_0)\Omega)}=1.\label{gml5}
\end{equation}
A possibility to ensure the validity of this equation is the above
assumption (which has not been used so far) that ${\cal L}_0$
and ${\cal L}_1$ have mass dimension ${\rm dim}({\cal L}_j)\leq 4$
and are treated as dimension $4$ vertices in the renormalization 
procedure. Due to this 
additional assumption and the requirements that the adiabatic limit
exists and is unique, the normalization of the vacuum diagrams is uniquely 
fixed, and with this normalization the vacuum diagrams vanish in the 
adiabatic limit
\begin{equation}
\lim_{\theta\to 1}(\Omega,S(\theta{\cal L}_0)\Omega)=1,\quad\quad
\lim_{\theta\to 1}(\Omega,S(\theta({\cal L}_0+\epsilon{\cal L}_1))\Omega)=1.
\end{equation}
(For a proof see also the appendix of \cite{D}.)  $\quad\w$

{\it Remarks}: (1) Without the assumption about ${\cal L}_0$
and ${\cal L}_1$ we find
\begin{equation}
\lim_{\theta\to 1}\Bigl(\Omega,S_{\theta ({\cal L}_0+\epsilon {\cal L}_1)}
(\lambda f)\Omega\Bigr)=\lim_{\theta\to 1}\frac
{\Bigl(\Omega,S_{\theta {\cal L}_0}(\theta\epsilon {\cal L}_1+\lambda f)
\Omega\Bigr)}{\Bigl(\Omega,S_{\theta {\cal L}_0}(\theta\epsilon {\cal L}_1)
\Omega\Bigr)}\label{Q7}
\end{equation}
instead of (\ref{Q6}), by using (\ref{Q4}) only. This is a formulation
of the QAP for general situations in which (\ref{gml5}) does not hold.

(2) By means of the QAP (\ref{Q2c}) (or (\ref{Q5}), or (\ref{Q7})) 
one can compute the change 
of the time ordered products of interacting fields (or of the 
Green's functions) under the variation of parameters $\lambda_1,...,\lambda_s$
if the interaction Lagrangian has the form ${\cal L}(x)=\sum_ia_i(\lambda_1,
...,\lambda_s){\cal L}_i(x),\>{\cal L}_i\in {\cal V}$
resp. ${\cal D}(\RR^4,{\cal V})$ (cf. eqns. (2.7-8)
of \cite{L}). But only the interaction ${\cal L}$ may 
depend on the parameters and not the time ordering operator (i.e.
the normalization conditions for the time ordered products). 

\section*{Appendix: wavefront sets and the pointwise pro\-duct 
of distributions}

In this appendix we briefly recall the definition of the 
wavefront set of a distribution
and mention a simple criterion for the existence of the pointwise product of 
distributions in terms of their wavefront sets.
For a detailed treatment we refer to 
H\"ormander \cite{H}, the application to quantum field theory
on curved spacetimes can be found in \cite{R,BFK,BF}.

Let $t\in {\cal D}'(\RR^d)$ be singular at the point $x$ and let
$f\in {\cal D}(\RR^4)$ with $f(x)\not= 0$. Then $ft\in {\cal D}'
(\RR^d)$ is also singular at $x$ and $ft$ has compact support.
Hence the Fourier transform $\widehat{ft}$ is a ${\cal C}^\infty$-function.
In some directions $\widehat{ft}$ does not rapidly decay, because otherwise
$ft$ would be infinitly differentiable at $x$. Thereby a function $g$
is called rapidly decaying in the direction $k\in \RR^d\setminus\{0\}$,
if there is an open cone $C$ with $k\in C$ and ${\rm sup}_{k'\in C}\>
|k'|^N|g(k')|<\infty$ for all $N\in {\bf N}$.

{\bf Definition:} The wavefront set ${\rm WF}(t)$ of a distribution
$t\in {\cal D}'(\RR^d)$ is the set of all pairs $(x,k)\in \RR^d
\times \RR^d\setminus\{0\}$ such that the Fourier transform
$\widehat{ft}$ does not rapidly decay in the direction $k$ for all
$f\in {\cal D}(\RR^d)$ with $f(x)\not= 0$.

For example the delta distribution satisfies $\widehat{f\delta}(k)=f(0)$,
hence ${\rm WF}(\delta)=\{0\}\times \RR^d\setminus\{0\}$.
The wavefront set is a refinement of the singular support of $t$ (which is
the complement of the largest open set where $t$ is smooth):
\begin{equation}
t\> {\rm is\> singular\>at}\>x\quad\Longleftrightarrow\quad\exists k\in
\RR^d\setminus\{0\}\>{\rm with}\>(x,k)\in {\rm WF}(t).\nonumber
\end{equation}
For the wavefront set of the two-point function one finds 
\begin{equation}
{\rm WF}(\Delta_+)=\{(x,k)\>|\>x^2=0,\,k^2=0,\,x\| k,\,k_0>0\}.\label{A1}
\end{equation}

Let $t$ and $s$ be two distributions which are singular at the same point 
$x$. We localize
them by multiplying with $f\in {\cal D}(\RR^d)$ where $f(x)\not= 0$.
We assume that $(ft)$ and $(fs)$ have only one overlapping singularity, namely
at $x$. In general the pointwise product $(ft)(y)(fs)(y)$ does not exist. 
Heuristically this can be seen by the divergence of the convolution integral
$\int dk\,\widehat{(ft)}(p-k)\widehat{(fs)}(k)$. But this integral
converges if $k_1+k_2\not=0$ for all $k_1,k_2$ with $(x,k_1)\in {\rm WF}(t)$
and $(x,k_2)\in {\rm WF}(s)$. This makes plausible the following theorem:

{\bf Theorem:} Let $t,s\in {\cal D}'(\RR^d)$ with
\begin{equation}
\{(x,k_1+k_2)\>|\>(x,k_1)\in {\rm WF}(t)\wedge (x,k_2)\in {\rm WF}(s)\}\>
\cap\>(\RR^d\times\{0\})=\emptyset.\label{A2}
\end{equation}
Then the pointwise product $(ts)\in {\cal D}'(\RR^d)$ exists.

By means of this theorem one verifies the existence of the distributional
products $(\varphi^{\otimes n})_\hbar (t)$ (\ref{W3}) and $(t\otimes_{k,
\hbar} s)$ (\ref{W7}).
\vskip0.5cm
{\bf Acknowledgements:} We thank Gudrun Pinter for several discussions on 
the quantum action principle, and Volker Schomerus  and Stefan Waldmann
for discussions on deformation quantization. In particular we are grateful 
to Stefan Waldmann for drawing our attention to reference \cite{Di}.


\begin{thebibliography}{999}

\bibitem{Bal}Balaban, T.,''Large Field Renormalization. II. Localization,
Exponentiation, and Bounds for the R Operation'',
{\it Commun. Math. Phys.} {\bf 122}, 355 (1989);
and earlier works of Balaban cited therein

\bibitem{BFFLS}Bayen, F.,Flato, M.,Fronsdal, C.,Lichnerowicz, A.,Sternheimer, 
D.,''Deformation Theory and Quantization'', {\it Annals of Physics (N.Y.)} 
{\bf 111}, 61, 111 (1978)

\bibitem{BRS} Becchi, C., Rouet, A., and Stora, R., "Renormalization of 
the abelian Higgs-Kibble model",
{\it Commun. Math. Phys.} {\bf 42}, 127 (1975)\\
Becchi, C., Rouet, A., and Stora, R., "Renormalization of gauge theories", 
{\it Annals of Physics (N.Y.)} {\bf 98}, 287 (1976)

\bibitem{BlSe}Blanchard, P., and S\'en\'eor, R., ``Green's functions for 
theories with massless particles (in perturbation theory)''
{\it Ann. Inst. H. Poincar\'e A} {\bf 23}, 147 (1975)

\bibitem{BDF}Boas, F.M., D\"utsch, M., and K.Fredenhagen, K.,"A local 
(perturbative) construction 
of observables in gauge theories: nonabelian gauge theories", work in progress

\bibitem{BS}Bogoliubov, N.N., and Shirkov, D.V., {\it "Introduction to the 
Theory of Quantized Fields"}, New York (1959)

\bibitem{BF}
Brunetti, R., and Fredenhagen, K.,
"Microlocal analysis and interacting quantum field 
theories: Renormalization on physical backgrounds", math-ph/9903028,
to appear in {\it Commun. Math. Phys.}

\bibitem{BFK}Brunetti, R., Fredenhagen, K., and K\"ohler, M.,''The microlocal
spectrum condition and Wick polynomials of free fields on curved space times'',
{\it Commun. Math. Phys.} {\bf 180}, 312 (1996)

\bibitem{DF}D\"utsch, M., and Fredenhagen, K.,
"A local (perturbative) construction of observables 
in gauge theories: the example of QED", 
{\it Commun. Math. Phys.} {\bf 203}, 71 (1999)

\bibitem{DF1}D\"utsch, M., and Fredenhagen, K.,
``Deformation stability of BRST-quantization'',
preprint: hep-th/9807215, DESY 98-098,
proceedings of the conference 'Particles, Fields and Gravitation', 
Lodz, Poland (1998)

\bibitem{Di}Dito, J., ``Star-Product Approach to Quantum Field Theory:
The Free Scalar Field'', {\it Lett. Math. Phys.} {\bf 20}, 125 (1990)

Dito, J., ``Star-products and nonstandard quantization for
Klein-Gordon equation'', {\it J. Math. Phys.} {\bf 33}, 791 (1992)

\bibitem{D}D\"utsch, M., ``Slavnov-Taylor identities from the causal point 
of view'', {\it Int. J. Mod. Phys. A} {\bf 12} 3205 (1997)

\bibitem{E} Epstein, H., ``On the Borchers' class of a free field'',
{\it N. Cimento} {\bf 27}, 886 (1963)

Schroer, B., unpublished preprint (1963)

\bibitem{EG}Epstein, H., and Glaser, V., "The role of locality in 
perturbation theory", {\it Ann. Inst. H. Poincar\'e A} {\bf 19}, 211 (1973)

\bibitem{H}H\"ormander, L.,The Analysis of Linear Partial Differential
Operators I'', Springer-Verlag, Berlin (1983)

\bibitem{IA}Il'in, V.A., Slavnov, D.A.,''Algebras of observables in the 
$S$-matrix approach'', {\it Theor. Math. Phys.} {\bf 36}, 578 (1978)

\bibitem{KO}Kugo, T., and Ojima, I., "Local covariant operator formalism of 
non-abelian gauge theories and quark confinement problem",
{\it Suppl. Progr. Theor. Phys.} {\bf 66}, 1 (1979)

\bibitem{Lam}Lam, Y.-M.P., ``Perturbation Lagrangian Theory for Scalar
Fields - Ward-Takahashi Identity and Current Algebra'', {\it Phys. Rev.}
{\bf D6} 2145 (1972); ``Equivalence Theorem on 
Bogoliubov-Parasiuk-Hepp-Zimmermann - Renormalized Lagrangian Field Theories'',
{\it Phys. Rev.} {\bf D7} 2943 (1973)

\bibitem{LSZ}Lehmann, H., Symanzik, K., Zimmermann, W.,''Zur Formulierung
quantisierter Feldtheorien'', {\it Nuovo Cimento} {\bf 1}, 205 (1955) 

\bibitem{L}Lowenstein, J.H., ``Differential vertex operations in Lagrangian 
field theory'', {\it Commun. Math. Phys.} {\bf 24}, 1 (1971)

\bibitem{MRS}Magnen, J., Rivasseau and S\'en\'eor, R.,''Construction of 
$YM_4$ with an infrared cutoff'', 
{\it Commun. Math. Phys.} {\bf 155}, 325 (1993)

\bibitem{PS}Piguet, O., and Sorella, S.P., {\it "Algebraic Renormalization''},
Springer-Verlag (1995)

\bibitem{Pi}Pinter, G.,{\it ``The Action Principle in Epstein Glaser
Renormalization and Renormalization of the $S$-Matrix of $\Phi^4$-Theory},
hep-th/9911063

\bibitem{P}Prange, D.,''Epstein-Glaser renormalization and differential 
renormalization'', {\it J. Phys. A} {\bf 32}, 2225 (1999)

\bibitem{R}Radzikowski, M.,''Micro-local approach to the Hadamard condition 
in quantum field theory on curved space-time'', {\it Commun. Math. Phys.} 
{\bf 179}, 529 (1996)

\bibitem{S}Scharf, G.,
{\it "Finite Quantum Electrodynamics. The causal approach"}, 
2nd. ed., Springer-Verlag (1995)

\bibitem{SD} Steinmann, O., ``Perturbation expansions in axiomatic field 
theory'', Lecture Notes in Physics {\bf 11}, Berlin-Heidelberg-New York:
Springer-Verlag (1971)

\bibitem{St2}Stora, R.,
"Differential algebras in Lagrangean field theory", 
ETH-Z\"urich Lectures, January-February 1993;\\
Popineau, G., and Stora, R.,
"A pedagogical remark on the main theorem of perturbative 
renormalization theory", unpublished preprint (1982)

\bibitem{Z}Zimmermann, W., in ``Lectures on Elementary Particles and 
Quantum Field Theory'', Brandeis Summer Institute in Theoretical Physics 
(1970), edited by S. Deser

\end{thebibliography}
\end{document}